\documentclass[twocolumn,prd,longbibliography,aps,superscriptaddress]{revtex4-1}

\usepackage{amsfonts,amsthm,amsmath,amssymb, upgreek,tipa,
  epsfig, color,  inconsolata,framed, mathtools, sidecap, graphics,
  graphicx, wrapfig, overpic,  mathrsfs, textcomp, verbatim,
  dsfont, bbold, 
  subfigure}
\pdfoutput=1
\usepackage[english]{babel}
\usepackage{todonotes,color}

\usepackage[utf8]{inputenc}
\usepackage[T1]{fontenc}
\usepackage{hhline}
\let\chapter\section
\usepackage[ruled,vlined]{algorithm2e}
\SetKwComment{comment}{\%}{}

\graphicspath{{Paper_Figures_LSDD/}{../Paper_Figures_LSDD/}}
\makeatletter
\def\input@path{{Paper_Figures_LSDD/}}
\makeatother

\date{\today}
\newcommand{\ee}{\mathrm{e}}
\newcommand{\len}{l}

\newcommand{\newu}{U}
\newcommand{\vect}[1]{\boldsymbol{#1}}

\newcommand{\eps}{\epsilon}

\newcommand{\de}{\delta}
\newcommand{\De}{\Delta}

\newcommand{\p}{\partial}
\newcommand{\dd}{\text{d}}

\newcommand{\xv}{\vect{x}}
\newcommand{\yv}{\vect{y}}
\newcommand{\vv}{\vect{v}}

\newcommand{\rj}{\mathcal{J}}

\newcommand{\NewPh}{\mathcal{P}}

\newcommand{\FCon}{\mathcal{C}}
\newcommand{\FDel}{\mathcal{D}}

\newcommand{\shSUM}[2]{\sum_{ K_{#1,#2} } }

\newcommand{\shSUMk}[2]{\sum_{ \kv_{#2} \in K_{#1,#2} } }
\newcommand{\shSUMAdjA}[2]{\sum_{ \Ad_{n} \in S_{#1} } }

\newcommand{\sv}{\vect{s}}
\newcommand{\Sv}{\vect{S}}
\newcommand{\vs}{\vec{s}}
\newcommand{\mS}{\mathds{S}}

\newcommand{\kv}{\vec{k}}

\newcommand{\FF}{\mathcal{F}}
\newcommand{\FAd}{\mathscr{F}}

\newcommand{\Ad}{A}

\newcommand{\KS}{\rho_{\text{KS}}}

\newcommand{\vx}{\vec{x}}

\newcommand{\LO}{\mathcal{L}}

\newcommand{\shA}{b}
\newcommand{\shB}{a}

\newcommand{\paulF}{f}

\newcommand{\Min}{m}
\newcommand{\Ob}{\rho_{\text{Ob}}}
\newcommand{\CMin}{C_\Min}

\newcommand{\ExcrCy}{r_{\text{exc}}^{(\text{Cy})} }
\newcommand{\ExcrOb}{r_{\text{exc}}^{(\text{Ob})} }
\newcommand{\Db}{D_{\text{burial}}}
\newcommand{\Kf}{\kappa_{\text{form}}}
\newcommand{\Om}{\Omega}

\begin{document}

\title{A Mean-Field Approach to Evolving Spatial Networks, with an Application to Osteocyte Network Formation}

\author{Jake P.~Taylor-King}
\affiliation{Mathematical Institute, University of Oxford, Oxford, OX2 6GG, UK}
\affiliation{Department of Integrated Mathematical Oncology, H. Lee Moffitt Cancer Center and Research Institute, Tampa, FL, USA}

\author{David Basanta}
\affiliation{Department of Integrated Mathematical Oncology, H. Lee Moffitt Cancer Center and Research Institute, Tampa, FL, USA}

\author{S.~Jonathan Chapman}
\affiliation{Mathematical Institute, University of Oxford, Oxford, OX2 6GG, UK}

\author{Mason A.~Porter}
\affiliation{Department of Mathematics, University of California Los Angeles, Los Angeles, 90095, USA}
\affiliation{Mathematical Institute, University of Oxford, Oxford, OX2 6GG, UK}
\affiliation{CABDyN Complexity Centre, University of Oxford, Oxford, OX1 1HP, UK}

\date{\today}

\begin{abstract}

We consider evolving networks in which each node can have various associated properties (a state) in addition to those that arise from network structure. For example, each node can have a spatial location
and a velocity, or some more abstract internal property that describes
something like social trait. Edges between nodes are created and destroyed, and new
nodes enter the system. We introduce a ``local state
degree distribution'' (LSDD) as the degree distribution at
a particular point in state space. We then make a mean-field assumption and thereby derive an integro-partial differential
equation that is satisfied by the LSDD. We perform numerical experiments and find good agreement between solutions of the integro-differential equation and the LSDD from stochastic simulations of the full model. To illustrate our theory, we apply it to a simple continuum model for osteocyte network formation within bones, with a view to understanding changes
that may take place during cancer. Our results suggest that 
increased rates of differentiation lead to higher densities of
osteocytes but with a lower number of dendrites.
To help provide biological context, we also include an introduction to osteocytes, the formation of osteocyte networks, and the role of osteocytes in bona metastasis.

\end{abstract}

\maketitle

%%%%%%%%%

%%%%%%%%%

\section{Introduction}\label{sec_intro}

Networks, in which entities (``nodes'') interact with each other via
``edges'', are a useful representation of complex systems \cite{Newman_2010}. They have often been very helpful for formulating and
answering questions in biology, sociology, engineering, and numerous other fields. Because the present work is motivated by a biological application, let's consider a few examples from biology. Many systems---such as
blood vasculature \cite{Jain_1988}, leaf venation \cite{Blonder_2011},
and fungi \cite{heaton2010,shl2015}---can be treated as biological
transportation networks, in which edges carry resources and nodes
operate as junctions.  Some of these studies exploit ideas from fluid mechanics and energy
minimization to investigate flow through various media
\cite{Liesbet_2013,Hu_2013,Haskovec_2015,heaton2010}. To give another type of example, in evolutionary
game theory, a node can represent a biological agent, and edges
indicate interactions in a ``game'' between those agents
\cite{Szabo_2007}. Applications range from behavioral ecology
\cite{Smith_1982} to investigating tumor heterogeneity in cancer
\cite{Archetti_2013}. Changes in node fitness can depend on, for example, a node's
phenotype and its immediate neighbors \cite{Ohtsuki_2006}.  

Many of the above examples involve ``spatial networks''
\cite{Barthelemy_2011}, and spatial constraints can exert significant
influence (directly and/or indirectly) on both network structure and function. In the
aforementioned examples, the networks are embedded in space, and one
thus can assign physical locations to the nodes and edges. This is
clearly important when considering dynamical processes on those
networks \cite{Durrett_1994}.  

Some spatial networks grow in time as they form: new nodes and
edges can join a spatial network, and the spatial domain can
expand. For example, cities often grow outwards or arise when borders
from multiple settlements coalesce \cite{Gallotti_2014, Louf_2014, barth2017}. Fungi, which are living networks, expand to reach nutrients, and such growth induces flows of mass \cite{heaton2010,shl2015}.

In the present paper, we propose a
framework for describing evolving spatial networks. We then apply
this framework to examine the formation of osteocyte networks in
bone (see Fig.~\ref{fig_bone_laying_diagram} for a schematic). During bone formation, cells called osteoblasts secrete bone matrix and differentiate into cells called osteocytes, and the ensuing growth
process results in a network of connected cells that communicate by
chemical diffusion via gap junctions \cite{Buenzli_2015, Dallas_2010,
  Franz-Odendaal_2006, Ishihara_2012}.  In Section \ref{sec_first_model_osteo}, we develop and analyze a model for this process, with motivation of using network analysis to study bone cancer. In pathological bone, the highly regulated bone-remodeling signaling pathway is disrupted, and it may be possible to gain insight into the nature of this disruption using tools from network analysis.
  
There are myriad models of network formation \cite{Newman_2010}. There
are at least three possible ways of formulating such a model:
(\emph{i}) all nodes and edges are created simultaneously with a
single algorithmic step (e.g., the standard Erd\H{o}s--R\'{e}nyi (ER)
random graph $G(n,p)$ \cite{Erdos_1960,Newman_2010} and standard
random geometric graphs \cite{penrosergg}); (\emph{ii}) nodes and
edges have an implicit order of creation but time is not considered
explicitly (e.g., in some preferential-attachment models
\cite{Price_1986,Moore_2006}); or (\emph{iii}) nodes and edges have an
order of creation and time is considered explicitly (e.g., in some
preferential-attachment models \cite{Krapivsky_2001, Krapivsky_2001_b}
and in adaptive network models \cite{thilo-adapt2}). Because we want to incorporate time explicitly, we will consider
spatial networks in category (\emph{iii}) in the present paper. See
also the recent work by Zuev et al. on geometric
preferential-attachment models \cite{bianconi2015}. 

When studying a model in category (\emph{iii}), it is common to employ
kinetic approximations \cite{Krapivsky_2010, Krapivsky_2001,
  Krapivsky_2001_b, Marceau_2010}. Such approximations often allow one
to construct a ``master equation'' to obtain approximate and/or
asymptotic expressions for quantities such as degree distributions,
component sizes, and cycle sizes \cite{Krapivsky_2010}. These
equations can take the form of an ordinary differential equation
(ODE), partial differential equation (PDE), or other continuous
model. By carefully constructing a general state space of the system,
we derive an extension to the ODE master equations given in
\cite{Krapivsky_2010, Krapivsky_2001, Krapivsky_2001_b, Marceau_2010}
to obtain a master equation in the form of an integro-partial
differential equation (IPDE) that incorporates this state
space.  

Our work illustrates how to use a master-equation approach to study
spatial networks. As an illustration of its potential, we examine
degree distributions in a very general model of evolving spatial
networks and use our results to gain insights on osteocyte network
formation in bone. An important benefit of using an explicit
time-dependent kinetic approach is that it allows one to incorporate
nodes that move through a state space (e.g., particles that
diffuse). Note that we will often use the terms ``nodes'' and
``particles'' interchangeably (depending on the context). When
considering spatially-embedded network models, the state space
corresponds to each node being located in a copy of the physical
space. From a mathematical viewpoint, our approach is reminiscent of
some models of social networks \cite{Boguna_2003, Wong_2006}, for
which the state space corresponds to a latent social space described
by some internal parameters; however, these models have no time
dependence. See also the recent paper \cite{barre2017} on macroscopic descriptions for particle interactions mediated by time-dependent networks.

The remainder of our paper is organized as follows. In Section
\ref{sec_LSDD}, we define the concept of a ``local state degree
distribution'' (LSDD), which encapsulates the degree distribution of a
network local to a point in state space. In Section
\ref{sec_description}, we give a description of our model of evolving
spatial networks. In Section \ref{sec_edge_creation}, we derive an
equation for the LSDD when edges are created but cannot be
destroyed. In Section \ref{sec_edge_deletion}, we incorporate edge
destruction into our model. Our derivations require the use of
approximations, so we use numerical simulations to explore agreement and discrepancies between
theory and our derived equations in Section \ref{sec_numerics}. In Section \ref{sec_first_model_osteo}, we use
our model for evolving spatial networks to formulate a model for
osteocyte network formation. We conclude and discuss future directions
in Section \ref{conc}. In appendices, we give additional details on
derivations and numerical implementations. 

%%%%%%%%%%%%%%%%%%%%%%%%%%%%%%%%%%%%
\section{A Local State Degree Distribution}\label{sec_LSDD}
%%%%%%%%%%%%%%%%%%%%%%%%%%%%%%%%%%%%

We consider networks in which each node (i.e., particle) has some number of associated
properties in addition to those, such as degree distribution, that arise from network structure. For example, each node may have a spatial location and a velocity, or it may have some more abstract internal (``latent'') property describing, for example, some social
trait. We collect these properties together into a state vector $\sv$, which belongs to a state space $\mS$.

The density $\paulF (t,\sv)$ of particles in the state space gives the expected number of
particles with state $\sv$ \footnote{More precisely, $\paulF (t,\sv)\dd
  \sv$ gives the expected number of particles that have states lying in
  the volume element $\dd \sv$ centered on $\sv$.}. However, a common way to study the properties of models of
network formation is to examine the degree distribution \cite{Newman_2010}.  
In this paper, we combine these ideas to consider what we 
call a \emph{local state degree distribution} (LSDD) $u_k(t, \sv)$,
which gives the expected number of particles of degree $k$ at time $t$ with
state vector $\sv$. One can write the LSDD as
\begin{equation*}  
	u_k(t, \sv)=  p_k(t \,|\, \sv)\paulF (t,\sv)\,,
\end{equation*}
where
 $p_k(t \,|\, \sv )$  is the conditional probability that a node at time $t$
has degree $k$, given that 
its state is $\sv$. 
We can recover both  $\paulF$ and 
$p_k$ from $u_k$ via
\begin{equation}
	\paulF (t,\sv) =  \sum_{k=0}^\infty u_k(t, \sv)\,, \quad
        p_k(t \,|\, \sv) = \frac{u_k(t, \sv)}{\sum_{k=0}^\infty u_k(t,
          \sv)}\, . \label{fandpk}
\end{equation}
The degree distribution of the whole network is given by
\begin{equation}
P_k(t) = \frac{ \int_{\mS} u_k(t,
          \sv) \dd \sv }{\sum_{k=0}^\infty \int_{\mS} u_k(t,
          \sv) \dd \sv} \, .
\end{equation}

%%%%%%%%%%%%%%%%%%%%%%%%%%%%%%%%%%%%
\section{Model of Evolving Spatial Networks}\label{sec_description}
%%%%%%%%%%%%%%%%%%%%%%%%%%%%%%%%%%%%

We now present a model for evolving spatial networks. Our model has
three tunable features.
First, edges can be created and deleted. Second, new nodes can be
created (but we do not allow node deletion).
Finally, we specify a model (possibly depending on network structure) for the evolution of
the state $\sv$ of each node.

Suppose at time $t$ that there are $N(t)$ nodes with state vectors
$\sv_i$ and degrees $k_i$ (with $i \in \{1, \ldots,N\}$).
We also suppose that new edges are created between each pair of nodes as
independent Poisson processes, where $\FCon(\sv_i, k_i ,\sv_j, k_j)$ is the
rate of edge creation between node $i$ and node $j$, so that the
probability of an edge being created between node $i$ and node $j$ in
time $t$ to $t+\dd t$ is $\FCon(\sv_i, k_i ,\sv_j, k_j) \dd t$.
We also suppose that $\FCon$ depends on the states and degrees of the two
nodes $i$ and $j$, but that it does not depend on other properties of the network (such
as, for example, whether an edge between node $i$ and node $j$
already exists). Thus, our model allows \emph{multiedges} (i.e., multiple
edges between two distinct nodes).

Similarly, we suppose that existing nodes are deleted as independent
Poisson processes, where $\FDel(\sv_i,
k_i,\sv_j, k_j)$ is the rate of edge deletion (per edge) between node
$i$ and node $j$. We suppose also that new particles, which have degree $0$, arrive
randomly as a Poisson process with constant 
rate $\rj$, and we assign to them a state drawn at random from the
probability distribution $\NewPh$. 

The final component of our model is the equation of motion of the
particles in the state space. Many possible models are available (both
deterministic and stochastic), and we want to keep our presentation as
general as possible. Nevertheless, it is useful to have a model in
mind to fix ideas, and we thus present several examples.

The simplest case is for each node to have a constant state vector.
Our first nontrivial example consists of identical particles of mass $m>0$ that follow Newton's
equations of motion with a smooth pairwise potential $\Phi$. In that
case, 
$\sv_i(t) = (\xv_i(t),\vv_i(t))\in \mS \subseteq \mathds{R}^{2d}$ and 
\begin{equation}\label{eq_Newton_EoM}
	\vect{\dot{x}}_{i} = \vv_{i}\,, \quad m\vect{\dot{v}}_{i} = -
        \sum_{\substack{j=1 \\ j\neq i}}^{n} \nabla_{\xv_i}\Phi(\xv_i
        - \xv_j) \, . 
\end{equation} 
Variations of these equations have been used 
to consider collective motion such as swarming \cite{Degond_2004, Gallagher_2013, Illner_1987}. 
Our second example consists of the stochastic 
 position jump process in which
$\sv_i(t) = \vect{X}_i(t)\in\mS \subset \mathds{R}^d$ and
\footnote{Following standard practice, we use capital letters for
  random variables and lower-case letters for realizations of these variables.} 
\begin{equation}\label{eq_SDE_EoM}
	\dd \vect{X}_i = \vect{\mu}(\vect{X}_i)\dd t + \sigma\,\dd
        \vect{W}_t\,, 
\end{equation}
where $ \vect{W}_t$ is standard Brownian motion
(i.e., a Wiener process)
\cite{Oksendal_2014}. Other stochastic examples include velocity-jump processes
\cite{Othmer_1988} and fractional diffusion processes
\cite{Metzler_2000}. 

In each of the above examples, the motion in state space is independent of network structure. Of course, it
is also possible to imagine scenarios in which the 
motion depends on node degree or other structural features.

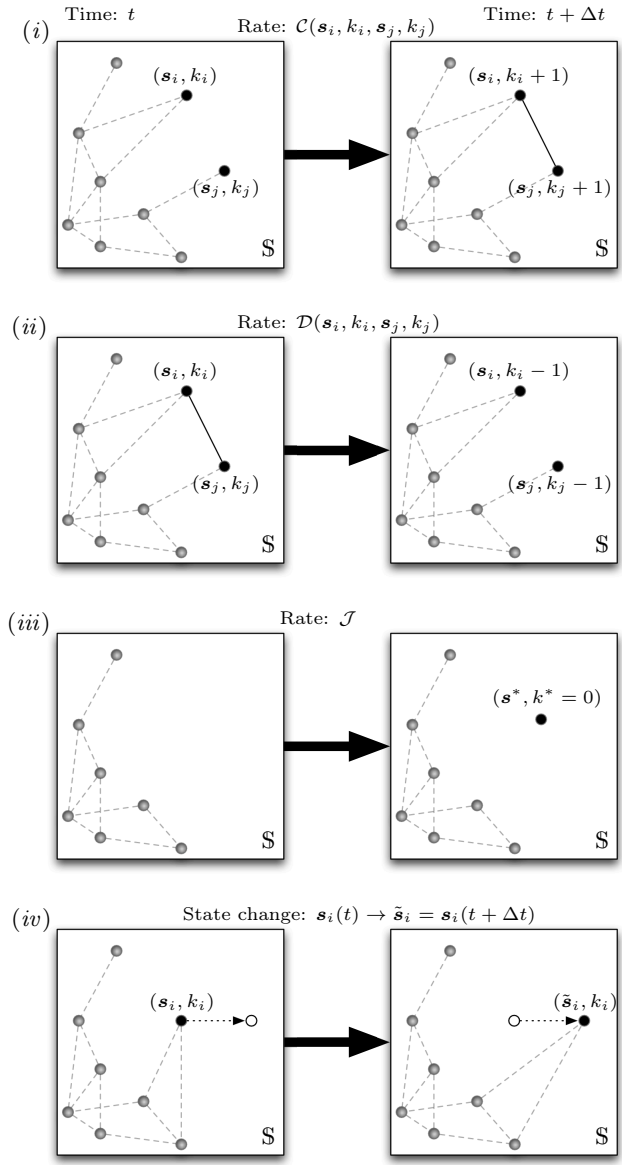
\begin{figure}[h!]
%\iffalse
   \begin{overpic}[width=0.45\textwidth]{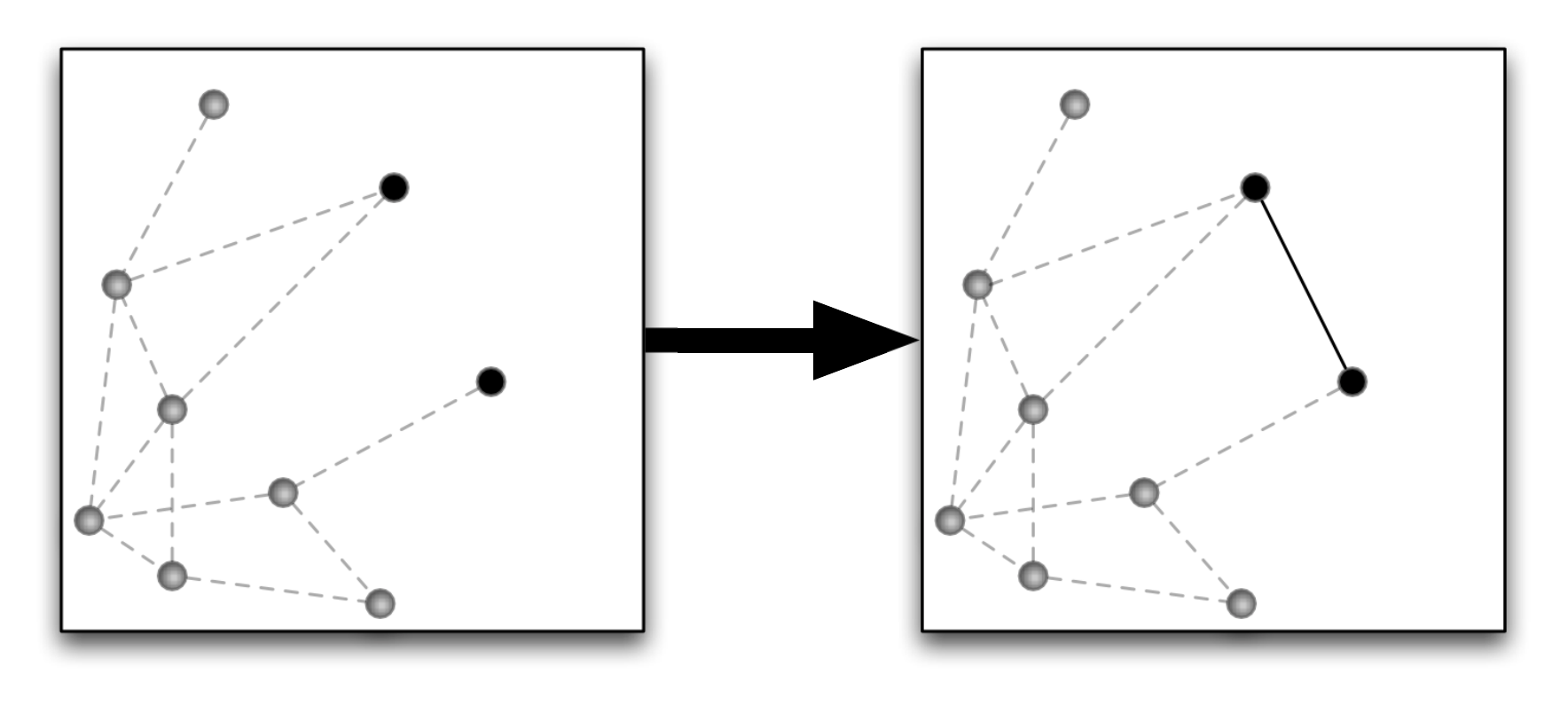} 
   	\put(-2,42.5){(\emph{i}) }
   	\put(5,45.5){\scriptsize Time: $t$}
	\put(74,45.5){\scriptsize Time: $t+\De t$}
	\put(33.5,43.5){\scriptsize Rate: $\FCon(\sv_i,k_i,\sv_j,k_j)$}
  	\put(19.5,35.5){\scriptsize $(\sv_i,k_i)$}
	\put(26,17){\scriptsize $(\sv_j,k_j)$}  
	\put(71.5,35.5){\scriptsize $(\sv_i,k_i+1)$}
	\put(78,17){\scriptsize $(\sv_j,k_j+1)$} 
	\put(37.5,6.5){$\mS$} 
	\put(92.5,6.5){$\mS$}  
  \end{overpic} \\ \vspace{10pt}
   \begin{overpic}[width=0.45\textwidth]{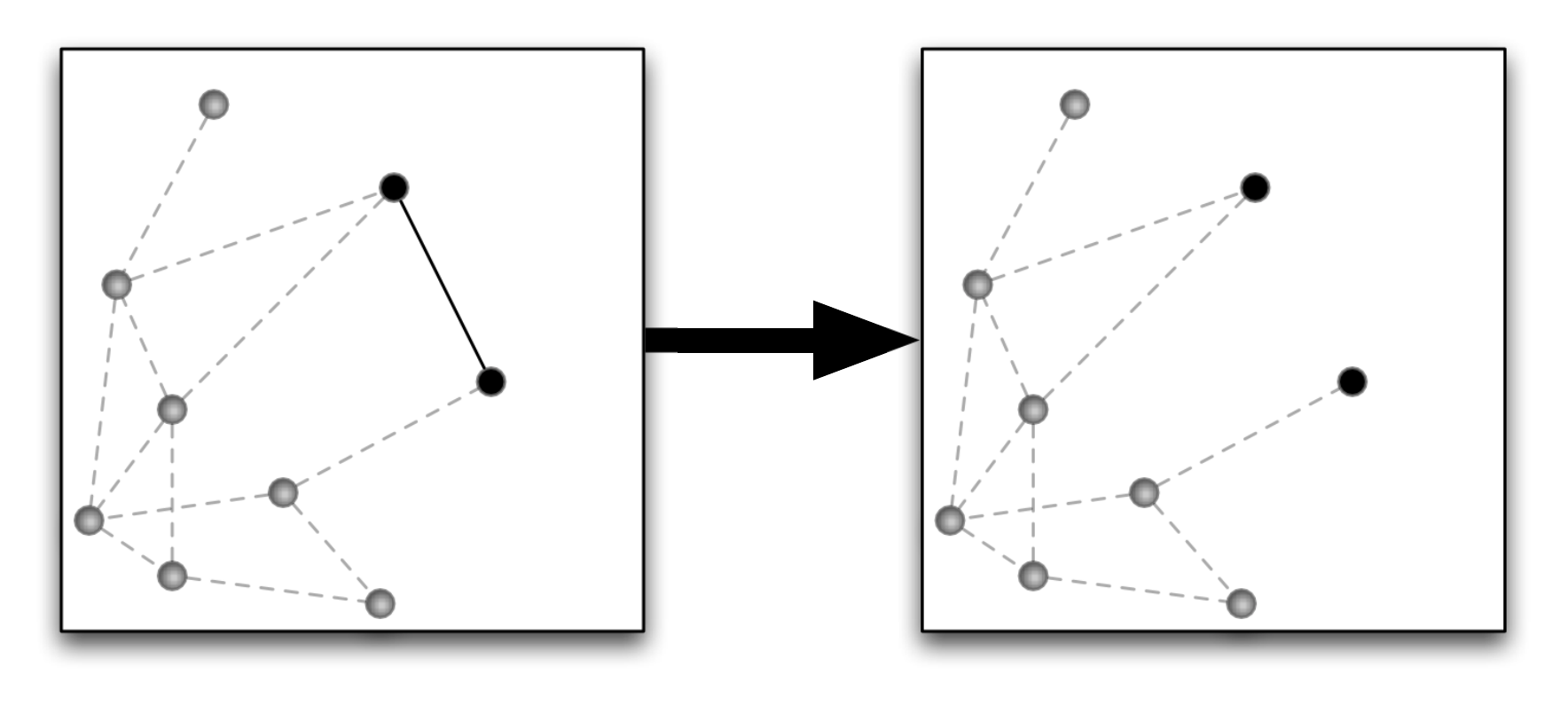} 
      	\put(-3.75,42.5){(\emph{ii}) }
   	\put(33.5,43.5){\scriptsize Rate: $\FDel(\sv_i,k_i,\sv_j,k_j)$}
  	\put(19.5,35.5){\scriptsize $(\sv_i,k_i)$}
	\put(26,17){\scriptsize $(\sv_j,k_j)$}  
	\put(71.5,35.5){\scriptsize $(\sv_i,k_i-1)$}
	\put(78,17){\scriptsize $(\sv_j,k_j-1)$}  
	\put(37.5,6.5){$\mS$} 
	\put(92.5,6.5){$\mS$}    
  \end{overpic}  \\ \vspace{10pt}
   \begin{overpic}[width=0.45\textwidth]{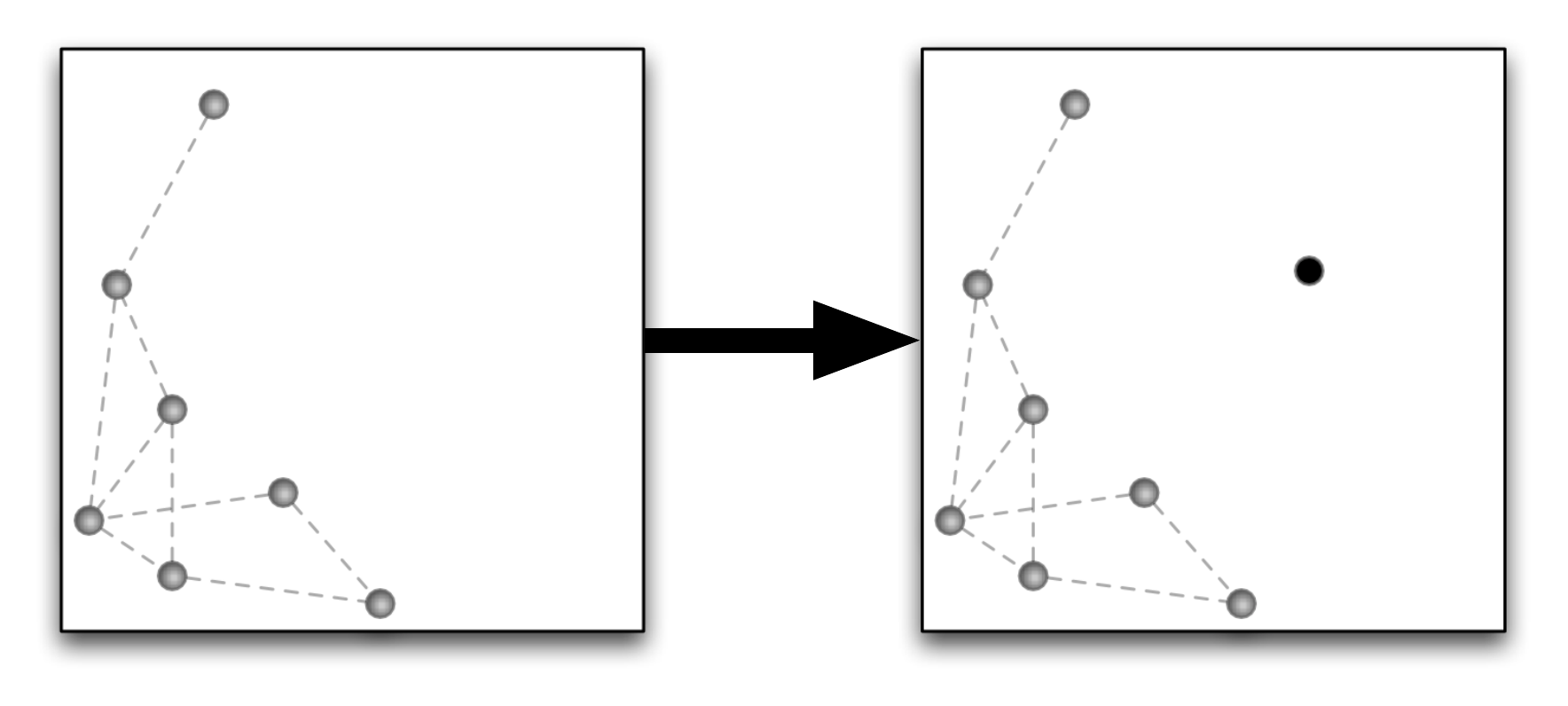} 
      	\put(-4.5,42.5){(\emph{iii}) }
      	\put(40.5,43.5){\scriptsize Rate: $\rj$}
	\put(75.5,30.5){\scriptsize $(\sv^*,k^*=0)$}  
	\put(37.5,6.5){$\mS$} 
	\put(92.5,6.5){$\mS$}  
  \end{overpic}  \\ \vspace{10pt}
   \begin{overpic}[width=0.45\textwidth]{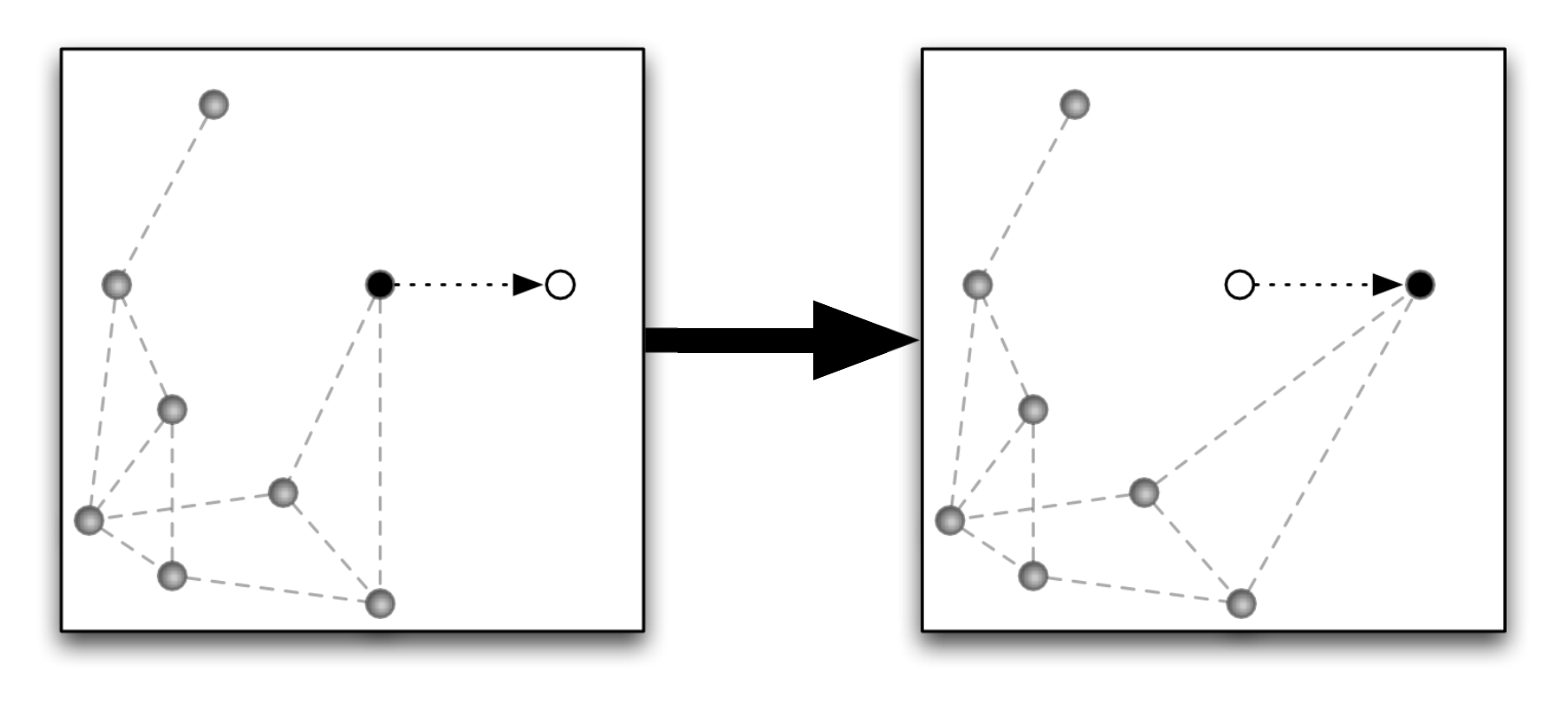} 
      	\put(-3.75,42.5){(\emph{iv}) }
         \put(24.5,43.5){\scriptsize State change: $\sv_i(t)\rightarrow \tilde{\sv}_i = \sv_i(t + \De t)$}
  	\put(19,29){\scriptsize $(\sv_i,k_i)$}  
  	\put(85.5,29){\scriptsize $(\tilde{\sv}_i,k_i)$}  
	\put(37.5,6.5){$\mS$} 
	\put(92.5,6.5){$\mS$}  
  \end{overpic}
%\fi
  \caption{
  \footnotesize{
  Diagrammatic illustration of our model of evolving spatial
  networks. In each panel, the square box represents the state space
  $\mS$. In each time step of size $\De t$ (where $0<\De t\ll 1$), the
  following events can occur: (\emph{i}) edge creation; (\emph{ii})
  edge deletion; (\emph{iii}) node creation; and (\emph{iv}) evolution
  of node state.}}
  \label{fig_algorithm_diagram}
\end{figure}

We summarize the model events
and state update in Table \ref{table_full_model}, and we illustrate
them in Fig.~\ref{fig_algorithm_diagram}.

%%%%%%%%%%%%%%%%%%%%%%%%%%%%%%%%%%%%

\begin{table}
%\begin{framed}
\caption{\footnotesize{Model description: (\emph{i}) edge creation;
    (\emph{ii}) edge deletion; (\emph{iii}) node creation; and
    (\emph{iv})  evolution of node state.}}
\hrulefill
\begin{itemize}

\item[(\emph{i})]
The rate of edge
  creation between nodes $i$ and $j$ is $\FCon(\sv_i, k_i, \sv_j,
k_j)$.  

\hrulefill

\item[(\emph{ii})]
The  rate of edge deletion per edge  between nodes $i$ and $j$ is
$\FDel(\sv_i, k_i, \sv_j, k_j)$.  

\hrulefill

\item[(\emph{iii})]
Nodes of degree $0$  enter the system at rate $\rj$. We assign the new node a state  $\sv^*\in\mS$, where we draw $\sv^*$ from the
distribution $\NewPh$. 

\hrulefill

\item[(\emph{iv})]
Nodes move in the state space $\mS$ according to some (possibly
stochastic) differential equation. We specify a single differential
equation for each node, and all nodes must follow the same
differential equation.   

\end{itemize}

\hrulefill

%\end{framed}
\label{table_full_model}
\end{table}
%%%%%%%%%%%%%%%%%%%%%%%%%%%%%%%%%%%%

%%%%%%%%%%%%%%%%%%%%%%%%%%%%%%%%%%%%
%%%%%%%%%%%%%%%%%%%%%%%%%%%%%%%%%%%%

Our model treats all nodes identically, although our methodology can
be extended to heterogeneous classes of nodes or heterogeneous classes of edges.
(Both of these generalizations are examples of multilayer networks
\cite{Kivela_2014}.) To consider such cases, one can modify kinetic
methodology from \cite{Bisi_2006, Burgers_1961}.

%%%%%%%%%%%%%%%%%%%%%%%%%%%%%%%%%%%%
\section{Model Analysis: No Edge Deletion}\label{sec_edge_creation}
%%%%%%%%%%%%%%%%%%%%%%%%%%%%%%%%%%%%

We begin by considering the case in which there is no edge
deletion. That is $\FDel(\sv_i,k_i,\sv_j,k_j)= 0$ for all
$\sv_i,k_i,\sv_j,k_j$. In the next subsection, we give a
set of hierarchical master Fokker--Planck (FP) equations for the
probability distribution of the state of the system. Because it is not pragmatic to work in this high-dimensional
space, in Section \ref{lowdim}, we reduce the dimension using
mean-field arguments from kinetic theory. 

% which leads to the derivation of a Boltzmann-like ``collision'' term. However, in this context, the collision is not a discontinuous jump in the velocity-space but a movement in the local state degree distribution.

%%%%%%%%%%%%%%%%%%%%%%%%%%%%%%%%%%%%
\subsection{Fokker--Planck Equation}\label{subsec_infinite_DE}
%%%%%%%%%%%%%%%%%%%%%%%%%%%%%%%%%%%%

We define $\FF_n^{\,\kv_n}(t, \vs_n)$ to be the
probability that a network has $n$ nodes with degree sequence $\kv_n
= \{k_1, \dots, k_n\}$ and state vectors $\vs_n = \{\sv_1, \dots,
\sv_n\}$. Note the normalization
\begin{equation}\label{eq_FF_normalisation}
	\sum_{n=0}^{\infty} \left\{ \shSUM{1}{n} \left[ \int_{\mS^n} \FF_n^{\kv_n}(t, \vs_n ) \, \dd \vs_n \right] \right\} = 1\,,
\end{equation} 
where $K_{a,b}$ is shorthand for summing over all possible degrees for nodes $i\in\{a,a+1,\dots,b\}$. That is,
\begin{equation}
 	\sum_{  K_{a,b}}\,\equiv\,\, \sum_{k_a = 0}^{\infty}\,  \sum_{k_{a+1} = 0}^{\infty} \dots  \sum_{k_b = 0}^{\infty}\, .
\end{equation}

We are not considering edge deletion and we have assumed
that edge creation and state-space motion depend only on node state and
 degree, so it is possible to write down a closed equation for
$\FF_n^{\,\kv_n}(t, \vs_n)$. In contrast, when we do consider edge deletion in 
Section \ref{sec_edge_deletion}, it will not be enough simply to keep
track of node degrees. We will need the full adjacency matrix.

Because the probability density function $\FF_n^{\,\kv_n}(t, \vs_n)$
depends on the number of particles in the system (which changes when new nodes are created), we
obtain a hierarchy of Fokker--Planck equations as in
\cite{Chen_2014}. By considering a small time step from $t$ to $t +
\dd t$ and partitioning over the events that can occur, we obtain
\begin{widetext}
\begin{align}
	\left(\frac{\p}{\p t} + \LO^{(n)} \right) \FF_n^{\,\kv_n}(t,
        \vs_n) =  \sum_{i=1}^n \sum_{j =i+1}^n \left( \FCon(\sv_i,
          k_i-1, \sv_j, k_j-1) \FF_n^{\,\kv_{n,-}^{i,j} }(t, \vs_n) -
          \FCon(\sv_i, k_i, \sv_j, k_j) \FF_n^{\,\kv_n}(t, \vs_n)
        \right) \nonumber \\ 
+ \sum_{i=1}^n \frac{1}{n} \de_{k_i, 0} \,\rj \,\NewPh(\sv_i)
  \FF_{n-1}^{\,\kv_{n}^{i-}}(t,  
\vs_{n}^{i-}) - \rj \FF_n^{\,\kv_n}(t, \vs_n) \,
. \quad\quad\quad \label{eq_HFP_l1}  
\end{align}
\end{widetext}

The operator on the left-hand side describes the evolution of
particles in state space, and it depends on the particular model that one chooses
for this evolution. For example, if nodes move according to equation (\ref{eq_Newton_EoM}), one has the Liouville flux term
\begin{equation}\label{eq_Liou_N_dim}
	\LO^{(n)} \FF_n^{\,\kv_n}=   \sum_{i=1}^n\left( \vv_{i}\cdot\nabla_{\xv_{i}} +  \frac{\vect{F}(\vx_n)}{m}\cdot\nabla_{\vv_{i}} \right) \FF_n^{\,\kv_n} \,,
\end{equation}
where $\vx_n = (\xv_1,\ldots,\xv_n)$ and 
\[ \vect{F}(\vx_n) =  -
        \sum_{\substack{j=1 \\ j\neq i}}^{n} \nabla_{\xv_i}\Phi(\xv_i
        - \xv_j) \, . \]
Alternatively, if $\sv_i(t)$ evolves according to the stochastic differential equation (SDE) 
(\ref{eq_SDE_EoM}), one has the Kolmogorov forward operator
\begin{equation}\label{eq_Kolm_forward_N_dim} 
	\LO^{(n)}\FF_n^{\,\kv_n} =   \sum_{i=1}^n\nabla_{\xv_i}\cdot
        \left( \vect{\mu}(\xv_i) - \frac{\sigma^2}{2} \nabla_{\xv_{i}}
        \right) \FF_n^{\,\kv_n}\,. 
\end{equation}
If the states of nodes are time-independent, then $\LO^{(n)}\equiv 0$. 

 The first term in parentheses on the right-hand side of equation \eqref{eq_HFP_l1}
 corresponds to edge-creation events
 between each pair, $i$ and $j$, of nodes. The
 positive term corresponds to gaining a network with degree sequence
 $\kv_n = \{k_1, \dots, k_n\}$ from a network with degree sequence $\kv_{n,-}^{i,j} = \{k_1,
\dots, k_i - 1, \ldots, k_j - 1, \dots, k_n\}$ by adding an edge
between nodes $i$ and $j$. The negative term
 corresponds to losing a network of degree sequence $\kv_n$ (as it changes
to a network of degree sequence $\kv_{n,+}^{i,j} = \{k_1,\dots, k_i + 1, \dots, k_j + 1, \dots,
 k_n\}$) by adding an edge between nodes $i$ and $j$.

The second term on the right-hand side of equation
 \eqref{eq_HFP_l1} corresponds gaining a network with degree sequence $\{k_1, \dots,k_{i-1},0,k_{i+1},\dots, k_{n}\}$ from a network with degree sequence
$\kv_{n}^{i-} = \{k_1, \dots,k_{i-1},k_{i+1},\dots, k_{n-1}\}$ by adding a new node (of degree $0$) to the
system. The Kronecker 
delta $\de_{k_i, 0}$ ensures that this term is present only when $k_i =
0$, corresponding to the new node having degree $0$. 
 One draws the new state $\sv_i$ from the probability distribution with density function $\NewPh$. (The
state vector of the existing nodes is $\vs_{n}^{i-}=(\sv_1,\dots,\sv_{i-1},\sv_{i+1},\dots,\sv_n)$.) 
We assign the label of the new node uniformly at random from the set
$\{1,\ldots,n\}$ (rather than assigning it to be the last node $n$) to
ensure that particles are indistinguishable. 

The final term on the right-hand side of equation
 \eqref{eq_HFP_l1} describes the loss of a network with degree sequence
 $\kv_{n}$ because of the addition of a new node. (One thereby obtains a network with degree
 sequence $\{k_1, \dots, k_{i-1},0,k_{i},\dots,k_{n}\}$ for some $i$.)

To save us from writing down separate equations for each case in which
$k_i=0$ for some $i$ (because it is impossible to arrive at a state in
which a node has degree $0$ by adding an edge to a state in which it
has degree $-1$), we use the convention that $\FF_n^{\kv_n}(t,\vs_n) = 0$
if $n<0$ or $k_i<0$ for all $i \in \{0,1,\dots,n\}$. 
We also suppose that particles are indistinguishable initially, so
the initial condition $\FF_n^{\,\kv_n}(0,\vs_n)$ is invariant to 
index permutation. Equation (\ref{eq_HFP_l1}) then ensures that this is true
for all $t$. 

It is not feasible to solve equation (\ref{eq_HFP_l1}) analytically
(except perhaps when the node state vectors are uncorrelated), and it
is not practical to solve it numerically due to the high dimension of
the domain. In Section \ref{lowdim}, we reduce the
dimension of the equation \eqref{eq_HFP_l1} using mean-field approaches from
kinetic theory.

%%%%%%%%%%%%%%%%%%%%%%%%%%%%%%%%%%%%
\subsection{Low-Dimensional Approximation}\label{lowdim}
%%%%%%%%%%%%%%%%%%%%%%%%%%%%%%%%%%%%

To derive our low-dimensional approximation, we adapt methods from
kinetic theory \cite{Carillo_2009, Degond_2004}. 
We keep the presentation brief in this subsection; we give more details in Appendix
\ref{App_finish_derivation_creation}. 

A common approach in
kinetic theory is to average over the states of particles $2$ to $n$ 
to find an equation for the marginal distribution function of the
first particle \cite{Cercignani_1988, Harris_1971} (the so-called ``1-particle
distribution function''). Because particles are indistinguishable, multiplying by $n$ gives the probability 
of finding {\em any} particle in a given state. Here we adopt the same
approach, and we average over the states and degrees of particles $2$ to
$n$. Because the number of particles itself can vary, we also need to
average over this quantity. The resulting 1-particle distribution function is exactly
the previously-defined LSDD $u_k(t,s)$. Specifically, 
\begin{widetext}
\begin{align}\label{above}
u_{k_1}(t,s_1) =   \sum_{n=0}^{\infty}  n\, \shSUM{2}{n} 
    \int_{\mS^{n-1} } \FF_n^{\kv_n}(t, \vs_n
    ) \, \dd \vs_n^{\, (2)}    
=  \sum_{n=0}^{\infty} n\, \sum_{k_{2}=0}^\infty \dots
\sum_{k_{n}=0}^\infty \int_{\mS^{n-1} }  \FF_n^{\kv_n}(t,
  \sv_1,\dots,\sv_n) \, \dd \sv_{2}\dots\dd \sv_n,
\end{align} 
\end{widetext}
where we introduce the shorthand notation
 $\dd \vs_n^{\,(\mu)} = \dd \sv_\mu \dots \dd \sv_n$ for $\mu \in \{1,\dots,n\}$.

To find the equation satisfied by $u_k(t,s_1)$, we apply the same
summation and integration to the Fokker--Planck equation
(\ref{eq_HFP_l1}).

%%%%%%%%%% part below is commented-out of pdf %%%%%%

\iffalse
The resulting equation for LSDD $u_{k_1}(t,\sv_1)$ (when no moment
closure are made) is the first equation in a BBGKY hierarchy. The
first equation then requires the evaluation of the 2-particle LSDD
$u^{(2)}_{k_1, k_2}(t,\sv_1,\sv_2)$. Furthermore, equation $k$ in the
BBGKY hierarchy requires the evaluation of the $(k+1)$-particle
LSDD. It is because of this that a moment closure argument is
required. In our case, after finding the first equation in the BBGKY
hierarchy, we use a mean field approximation.  
\fi

%%%%%%%%%

%%%% Time derivative
Because the summation and integration commutes with the time derivative, it follows for the first term on the left-hand side (LHS) of equation (\ref{eq_HFP_l1}) that
\begin{equation}\label{eq_low_dim_time_deriv}
	\sum_{n=0}^{\infty}n\, \shSUM{2}{n}  \int_{\mS^{n-1} }  \frac{\p \FF_n^{\kv_n}(t, \vs_n )}{\p t} \,\dd \vs_{n}^{\,(2)} = \frac{\p u_{k_1} (t, \sv_1)}{\p t} \, .
\end{equation}

%%%% LHS space integral

For the next term on the LHS, we need to evaluate 
\begin{equation}\label{eq_LO_eff_1}
	\sum_{n=0}^{\infty}n\,  \shSUM{2}{n}  \int_{\mS^{n-1} } 
        \LO^{(n)} \FF_n^{\kv_n} 
 \, \dd \vs_{n}^{\,(2)}   \, .
\end{equation}
If there are no interactions between particles in the state space, this
term evaluates to $\LO^{(1)}u_{k_1}$. Thus, for example, 
if $\LO^{(n)}$ is given by (\ref{eq_Kolm_forward_N_dim}), then this
term is 
\begin{equation}\label{eq_Kolm_forward_1_dim}
	\LO^{(1)}u_{k_1} =   \nabla_{\xv_1}\cdot \left( \vect{\mu}(\xv_1) - \frac{\sigma^2}{2}\nabla_{\xv_{1}} \right) u_{k_1}\,.
\end{equation}

When there are pairwise interactions between particles in the state space,
for each interacting pair, one can perform the integration over all
other particles. Consequently, after relabelling, one can write (\ref{eq_LO_eff_1}) in terms of the  
 2-particle LSDD 
\begin{equation}
	u^{(2)}_{k_1,k_2}(t,\sv_1,\sv_2) = \sum_{n=0}^{\infty}  n(n-1)\, \shSUM{3}{n} 
    \int_{\mS^{n-2} } \FF_n^{\kv_n}(t, \vs_n
    ) \, \dd \vs_n^{\, (3)}.\label{2particleULSDD}
\end{equation}
We do not have a closed equation for $u_{k_1}$, but the
first in a series of equations (the BBGKY hierarchy) for the
1-particle, 2-particle, 3-particle, etc. LSDDs. In this case, we make
the common mean-field closure assumption that 
\begin{equation}
	u_{k_1, k_2}^{(2)}(t, \sv_1, \sv_2) \approx u_{k_1} (t, \sv_1)
        u_{k_2} (t, \sv_2) \,. \label{eq_standard_molecular_chaos} 
\end{equation} 
For example, if $\LO^{(n)}$ is given by equation (\ref{eq_Liou_N_dim}), then
equation (\ref{eq_LO_eff_1}) becomes
\begin{equation}\label{eq_Boltzmann_operator}
	 \bar{\LO}^{(1)} u_{k_1} = \vv_1 \cdot  \nabla_{\xv_1} u_{k_1} - \mathcal{B}( \paulF,u_{k_1})\,,
\end{equation}
where $\mathcal{B}(\paulF,u_{k_1})$ is the mean-field approximation
\begin{equation}
	\mathcal{B}(\paulF,u_{k_1}) =  \left( \nabla_{\xv_1} \Phi * \int_{\mathds{R}^d} \paulF \,\dd \vv_1 \right)\cdot\left( \nabla_{\vv_1} u_{k_1} \right)\,,	\label{eq_Vlaslov_mean_field}
\end{equation}
where $*$ represents the convolution operator and the function $f$ is given in
equation (\ref{fandpk}).

%%%% RHS 
We now apply the same integration and summation to the right-hand side (RHS)
of equation 
\eqref{eq_HFP_l1}. We give a detailed derivation in Appendix 
\ref{App_finish_derivation_creation}. 
Here we simply note that by relabelling particles and again using the fact that 
$\FF_n^{\kv_n}(t, \vs_n )$ is invariant with respect to index permutation, we find that
\begin{widetext}
\begin{align}\label{eq_short_edge_coll}
\mbox{RHS} 
&=   \int_{\mS}   \sum_{{k_2} = 0}^{\infty} \left(  \FCon (\sv_1,
  k_1-1, \sv_2, k_2-1 )    u_{k_1-1,k_2-1}^{(2)} (t,  \sv_1,\sv_2)  -
  \FCon (\sv_1, k_1, \sv_2, k_2) u_{k_1,k_2}^{(2)} 
  (t,\sv_1, \sv_2) \right) \dd\sv_2 + \rj\NewPh ( \sv_1) \de_{k_1, 0}  \,.
\end{align} 
\end{widetext}
We again need to use the mean-field closure assumption
(\ref{eq_standard_molecular_chaos}) 
to write the 2-particle LSDD in terms of the 1-particle LSDD. This
gives the final closed mean-field equation for the 1-particle LSDD in
the absence of edge deletion:
%\newpage % If an error comes up when compiling, it is usually to do with use of {widetext}
\begin{widetext}
\begin{align}\label{eq_final_IPDE}
 	\left[ \frac{\p}{\p t} +  \LO^{(1)} \right] u_{k_1} (t,  \sv_1) &=  \left( \int_{\mS}   \sum_{{k_2} = 0}^{\infty}   \FCon (\sv_1, k_1-1, \sv_2, k_2-1 )    u_{k_2-1} (t,  \sv_2) \, \dd\sv_2 \right)  u_{k_1-1} (t, \sv_1) \nonumber  \\
&\quad -  \left(  \int_{\mS}\sum_{ {k_2} = 0}^{\infty} \FCon (\sv_1, k_1, \sv_2, k_2) u_{k_2} (t,  \sv_2) \, \dd\sv_2 \right) u_{k_1} (t, \sv_1) +  \rj\NewPh ( \sv_1) \de_{k_1, 0} \, ,
\end{align}
\end{widetext}
where $u_{k_1} \equiv 0$ if $k_1 <0$ by convention. 
When $\FCon$ is a constant and $\mS$ is a
point, equation \eqref{eq_final_IPDE} reduces to the master equations given in
Ref.~\cite{Krapivsky_2010}. The quadratic terms in equation (\ref{eq_final_IPDE}) are
analogous to the mean-field term in the Vlaslov 
equation, where a test particle feels the effect of a ``cloud'' of
points \cite{Cercignani_1988, Harris_1971, Gallagher_2013}.

%%%%%%%%%%%%%%%%%%%%%%%%%%%%%%%%%%%%
\section{Derivation of the Model: Edge Deletion}\label{sec_edge_deletion} 
%%%%%%%%%%%%%%%%%%%%%%%%%%%%%%%%%%%%

The state space $\{(\vs_n, \kv_n):n>0\}$ that we used in Section~\ref{sec_edge_creation}
  is not sufficient when we allow edge deletion. With edge deletion,
  it is crucial to know whether an edge exists between each pair of nodes,
  so we must consider the
  underlying adjacency matrix.
For undirected networks with
  multiedges, the adjacency matrix $\Ad_n = \Ad_n^T$ has entries $(\Ad_n)_{i,j}$
  for $i \neq j$ and $i,j \in \{1, \dots, n\}$, where 
  $(\Ad_n)_{i,j}\in\mathds{N}_0$ gives the number of edges between
  nodes $i$ and $j$. 

Because we consider probability distributions over $\Ad_n$, the most efficient representation is to restrict attention to the independent entries of $\Ad_n$. We thus change convention slightly and set $(\Ad_n)_{i,j}=0$ for $i \geq j$, and we will retain the term ``adjacency matrix'' to indicate the resulting matrix.

Let $\FAd_n^{A_n}(t,\vs_n)$ denote the probability that a network has adjacency matrix $\Ad_n$ and
$n$ nodes with state vectors $\vs_n = \{\sv_1,\dots,\sv_n\}$. The normalization condition is 
\small\begin{equation}
	\sum_{n=1}^\infty \left\{ \shSUMAdjA{n}{1} \left[ \int_{\mS^n}
            \FAd_n^{\Ad_n}(t,\vs_n)\, \dd\vs_n  \right]  \right\} =1\,
        , \label{deletenormal}
\end{equation} \normalsize
where 
\begin{equation}
	S_{n}  = \left\{\Ad_n: \begin{array}{ll}
	(A_n)_{ij} \in \mathds{N}_0\,, &  1\leq i < j \leq n \\[2mm]
	(A_n)_{ij} = 0\,, & \mbox{ otherwise}\end{array}
\right\} \, ,
\end{equation}
so that 
\begin{equation*}
	\shSUMAdjA{n}{1}\,\equiv\,\, \sum_{(\Ad_n)_{12}=0}^{\infty} \dots \sum_{(\Ad_n)_{1n}=0}^{\infty} \,\, \sum_{(\Ad_n)_{23}=0}^{\infty}  \dots  \sum_{(\Ad_n)_{n-1,n}=0}^{\infty} \, .
\end{equation*}
For fixed $n$, one can calculate the degree $k_i$ of node $i$ from the
adjacency matrix using 
\begin{equation*}
	k_i = \sum_{j=i+1}^n (\Ad_n)_{ij}+\sum_{j=1}^{i-1}(\Ad_n)_{ji}\,.
\end{equation*}	
Therefore, we can relate the distributions $\FF$ and $\FAd$ via 
\begin{widetext}
\begin{align}\label{eq_F_FAd_relation}
	\FF_n^{\kv_n}(t,\vs_n) &=  \shSUMAdjA{n}{1} \left[ \prod_{i=1}^n
                                 \de\left( k_i ,\, \sum_{j=i+1}^n
                                 (\Ad_n)_{ij}+\sum_{j=1}^{i-1}(\Ad_n)_{ji}
                                 \right)\right] \FAd_n^{\Ad_n}
                                 (t,\vs_n)\, , 
\end{align}
\end{widetext}
where $\de(a,b)$ is the Kronecker delta (which is usually written as $\de_{ab}$). When we include edge deletion, the hierarchical Fokker--Planck equation (\ref{eq_HFP_l1}) becomes
\begin{widetext}
\begin{align}
	\left(\frac{\p}{\p t} + \LO^{(n)} \right) \FAd_n^{ \Ad_{n} } (t, \vs_n) =  \sum_{i=1}^n \sum_{j =i+1}^n \FCon(\sv_i, k_i-1, \sv_j, k_j-1) \FAd_n^{ \Ad_{n,-}^{ij} } (t, \vs_n) - \FCon(\sv_i, k_i, \sv_j, k_j) \FAd_n^{ \Ad_{n} }  (t, \vs_n) \nonumber \\
+\sum_{i=1}^n \sum_{j=i+1}^n [(\Ad_n)_{ij} + 1] \FDel(\sv_i, k_i + 1, \sv_j, k_j + 1)  \FAd_n^{ \Ad_{n,+}^{ij} } (t,\vs_n) - (\Ad_{n})_{ij} \FDel(\sv_i, k_i , \sv_j, k_j )  \FAd_n^{ \Ad_{n} } (t,\vs_n)\nonumber \\
+ \frac{1}{n} \sum_{i=1}^n\left[\prod_{j=1}^{i-1} \de ({0,
  (\Ad_{n})_{ji}} ) \right]\left[\prod_{j=i+1}^{n} \de ({0, 
  (\Ad_{n})_{ij}} ) \right] \,\rj \NewPh(\sv_i) \FAd_{n-1}^{ \Ad_{n}^{i-}
  }(t, \vs_{n}^{\,i-})  - \rj \FAd_n^{ \Ad_{n} } (t, \vs_n) \label{eq_full_HFP} \, , 
\end{align}
\end{widetext}
where
\begin{equation}
	(\Ad_{n,\pm}^{ij})_{lm} = \left\{ \begin{array}{ll}
               (\Ad_n)_{lm} \pm 1\,,&
 \text{if $(i,j) = (l,m)$}\,, 
\\ (\Ad_n)_{lm}\,, 
& \text{otherwise\,.} \end{array} \right. 
\end{equation}
The first term on the RHS of equation \eqref{eq_full_HFP}
corresponds to edge-creation events between nodes $i$ and $j$ as
before (see Section \ref{sec_edge_creation}).
The first part of the second term corresponds to gaining a network with adjacency
matrix $\Ad_n$ from a network with adjacency matrix $\Ad_{n,+}^{ij}$ by deleting an edge
between $i$ and $j$. Note that $\FDel$ is the rate of deletion per
edge, so we multiply by the number of edges (which is equal to $(\Ad_n)_{ij}+1$) between $i$ and $j$. The second part of this term corresponds to losing a network with adjacency 
matrix $\Ad_n$ by deleting an edge
between $i$ and $j$ (to produce a network with adjacency matrix $\Ad_{n,-}^{ij}$).

The third term on the RHS of equation \eqref{eq_full_HFP} corresponds to gaining a network with adjacency 
matrix
\begin{widetext}
\begin{equation*}
	\left( \begin{array}{cccccccccc}
0 & A_{12} & \dots &A_{1,i-1} & 0 & A_{1,i+1} & A_{1,i+2} & \dots & A_{1n}\\
0 &0  & \dots &A_{2,i-1} & 0 & A_{2,i+1} &  A_{2,i+2} & \dots & A_{2n}\\
\vdots &&&&\vdots&&&&\vdots\\
0  & 0 & 0 & \dots & 0 &  A_{i-1,i+1} & A_{i-1,i+2} & \dots & A_{i-1,n}\\ 
0  & 0 & 0 &  \dots & 0 & 0 & 0&  \dots & 0\\
0 & 0 & 0 & \dots &  0 &  0 & A_{i+1,i+2} & \dots & A_{i+1,n}\\ 
0  & 0 & 0 & \dots & 0 & 0 & \dots & 0&A_{n-1,n} \\
0 & 0 & 0 & \dots & 0 & 0 & \dots &0& 0
	 \end{array}\right)
\end{equation*}
from an 
adjacency 
matrix 
\begin{equation*}
	\Ad_n^{i-} = \left( \begin{array}{cccccccc}
0 & A_{12} & \dots &A_{1,i-1}  & A_{1,i+1} & A_{1,i+2} & \dots & A_{1n}\\
0 &0  & \dots &A_{2,i-1}  & A_{2,i+1} &  A_{2,i+2} & \dots & A_{2n}\\
\vdots &&&&&&&\vdots\\
0  & 0 & 0 & \dots &  A_{i-1,i+1} & A_{i-1,i+2} & \dots & A_{i-1,n}\\ 
0 & 0 & 0 & \dots &    0 & A_{i+1,i+2} & \dots & A_{i+1,n}\\ 
0  & 0 & 0 & \dots & 0 & \dots & 0&A_{n-1,n} \\
0 & 0 & 0 & \dots & 0 & \dots &0& 0
 	\end{array}\right)
\end{equation*}
\end{widetext}
by relabeling nodes $j\rightarrow j+1$ for $j\geq i$ and
adding a new unconnected node with label $i$ (which we choose uniformly at random from the set
$\{1,\ldots,n\}$). The Kronecker $\delta$ 
ensures that this term is present only when $(\Ad_{n})_{ij}=(\Ad_{n})_{ji}=0$.
As in Section \ref{sec_edge_creation}, the uniformly random choice of
the label for the new node ensures that $\FAd_n^{ \Ad_{n} } (t, \vs_n)$ is invariant
with respect to index permutations.

We give the derivation of a reduced equation for $u_{k_1}(t,\sv_1)$ in Appendix 
\ref{App_finish_derivation_deletion}. When we include edge deletion, we
find that we need an additional closure assumption in addition to the mean-field
approximation \eqref{eq_standard_molecular_chaos}; see
Appendix \ref{App_finish_derivation_deletion} for details.
Our low-dimensional approximation to (\ref{eq_full_HFP}) is
\begin{widetext}
\begin{align}
	 \left[ \frac{\p}{\p t} +  \LO^{(1)} \right] u_{k_1} (t,
  \sv_1)  &=   u_{k_1-1} (t, \sv_1)  \left( \int_{\mS}   \sum_{{k_2} = 0}^{\infty}   \FCon
            (\sv_1, k_1-1, \sv_2, k_2-1 )    u_{k_2-1} (t,  \sv_2) \,
            \dd\sv_2 \right)  \nonumber \\ 
&\, -  u_{k_1} (t,
  \sv_1) \left(  \int_{\mS}\sum_{ {k_2} = 0}^{\infty} \FCon (\sv_1,
  k_1, \sv_2, k_2) u_{k_2} (t,  \sv_2) \, \dd\sv_2 \right)   \nonumber \\  
&+ 
\frac{(k_1 + 1)u_{k_1+1} (t, \sv_1)}{\int_{\mS}
  \sum_{k_1=0}^\infty k_1 
  u_{k_1}(t,\sv_1)\, \dd \sv_1} \left(  \int_{\mS}\sum_{ {k_2} =
  0}^{\infty} \FDel (\sv_1, k_1+1, \sv_2, k_2+1) (k_2+1)  u_{k_2+1}
  (t,  \sv_2) \, \dd \sv_2 \right)  \label{eq_del_addition} \\ 
&\, -    \frac{k_1  u_{k_1} (t, \sv_1)}{\int_{\mS}
  \sum_{k_1=0}^\infty k_1 
  u_{k_1}(t,\sv_1)\, \dd \sv_1} \left(  \int_{\mS}\sum_{ {k_2} =
  0}^{\infty}  \FDel (\sv_1, k_1, \sv_2, k_2)k_2 u_{k_2} (t,  \sv_2)
  \, \dd\sv_2 \right)  
%\nonumber \\  &
+ \rj\NewPh ( \sv_1) \de_{k_1, 0} \nonumber \, .
\end{align}
\end{widetext}

%%%%%%%%%%%%%%%%%%%%%%%%%%%%%%%%%%%%
\section{Numerical Examples}\label{sec_numerics}
%%%%%%%%%%%%%%%%%%%%%%%%%%%%%%%%%%%%

We now carry out Monte-Carlo simulations of our stochastic network
evolution process to illustrate the validity of equations
(\ref{eq_final_IPDE}) and (\ref{eq_del_addition}) for sufficiently
large networks.

First, in Section \ref{molchaos}, we investigate numerically 
the validity of the mean-field assumption for our model when the state space is a point.
Second, in Section \ref{sec_mock_scenario}, we use an example scenario to demonstrate that a numerical solution
  of equations (\ref{eq_final_IPDE}) and (\ref{eq_del_addition}) matches well with a full
Monte-Carlo simulation of the underlying process. Third, in Section 
  \ref{sec_Boltzmann_param}, we adapt the example scenario to consider
  the convergence of a network's degree distribution in the limit of
  large networks. Fourth, Section \ref{sec_one_step_kinetic}, we show that one can use our kinetic
approximation as an alternative to some one-step network
creation models proposed by Bogu\~{n}\'{a} \emph{et al.}
\cite{Boguna_2003}. Finally, in Section 
\ref{sec_numerics_finishing_remarks}, we briefly consider some
further approximations that one can make to equations (\ref{eq_final_IPDE}) and
(\ref{eq_del_addition}).

%%%%%%%%%%%%%%%%%%%%%%%%%%%%%%%%%%%%
\subsection{Accuracy of Our Mean-Field Assumption}\label{molchaos}
%%%%%%%%%%%%%%%%%%%%%%%%%%%%%%%%%%%%

In the absence of a network structure, the validity of the
mean-field approximation depends on the choice of $\LO^{(n)}$, and
it has been studied widely
\cite{Cercignani_1988, Harris_1971,Gallagher_2013,Bruna_2012,Degond_2004}. In the present paper, we focus on evolving network structure. In our exploration of the validity of the mean-field approximation in the network evolution model, we suppose that there is no state dependence in either the node creation rate $\FCon$ or the node deletion rate $\FDel$.

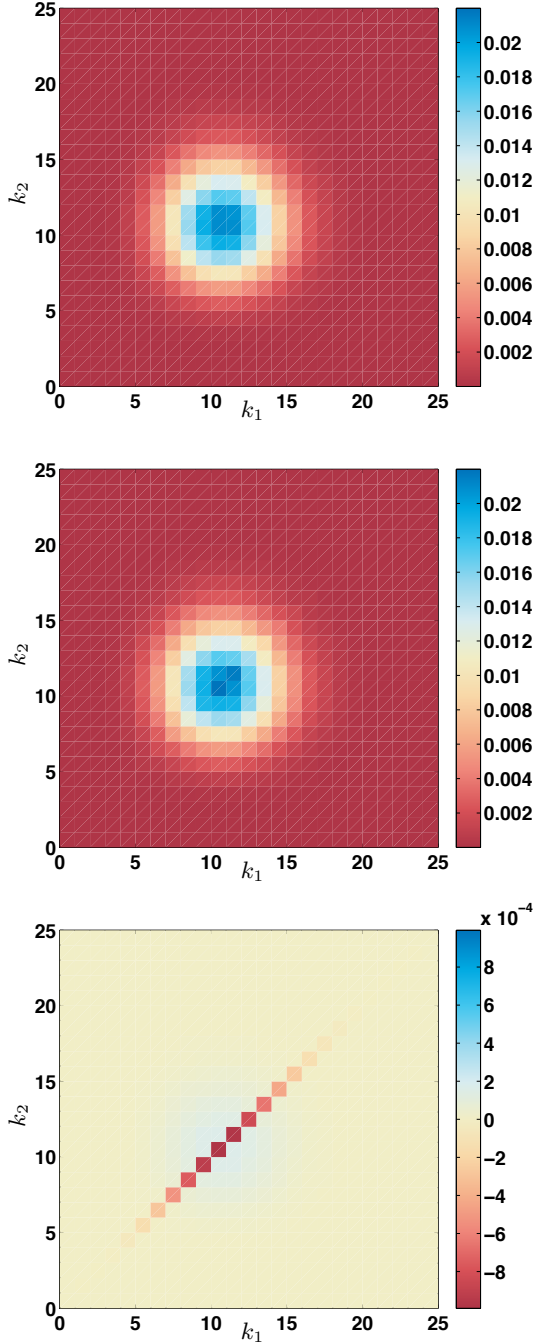
\begin{figure}[h!]
   \begin{overpic}[width=0.42\textwidth]{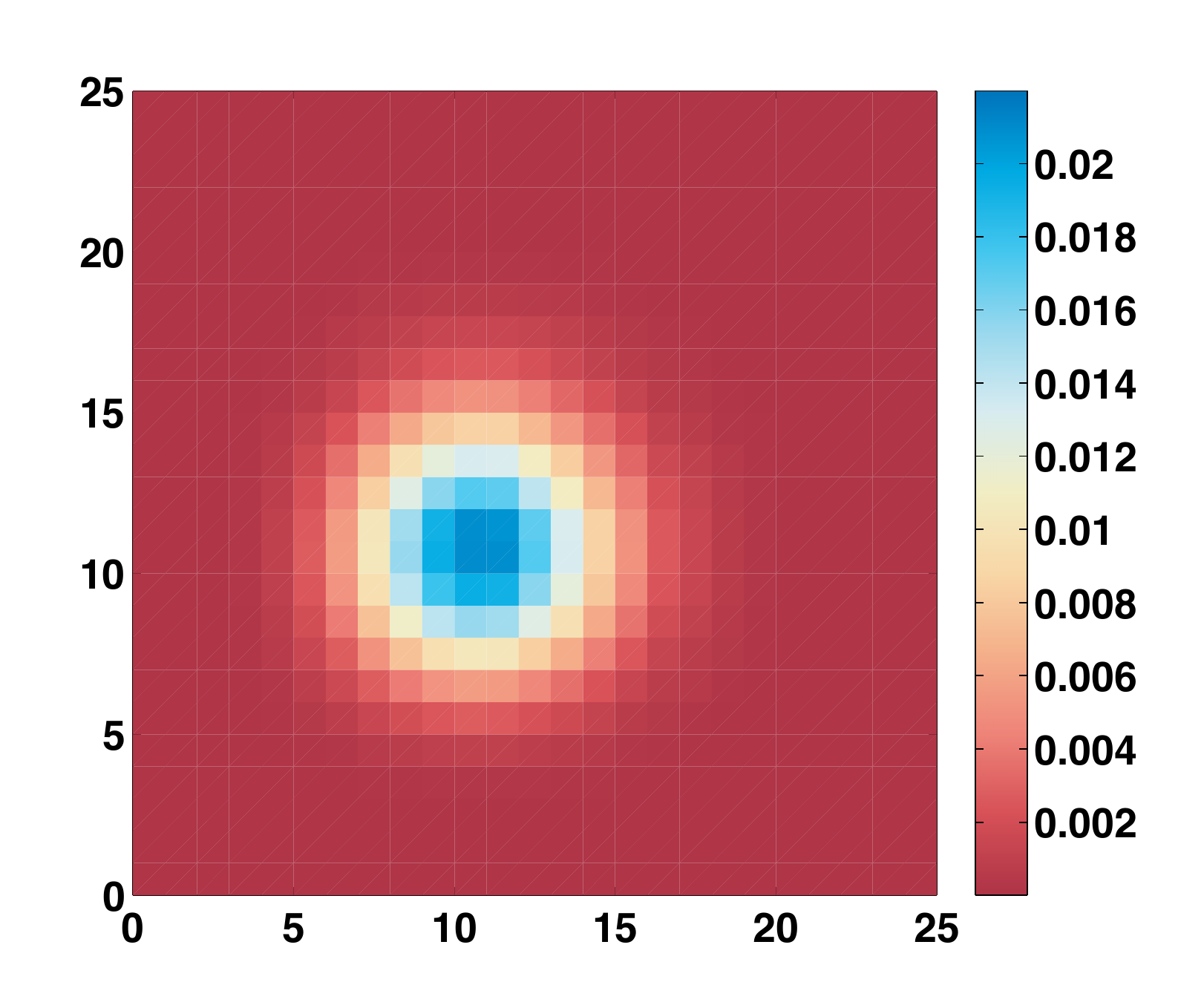} 
  	\put(2,39.5){\rotatebox{90}{$k_2$}}
	\put(43,2){$k_1$}  
  \end{overpic}
    \begin{overpic}[width=0.42\textwidth]{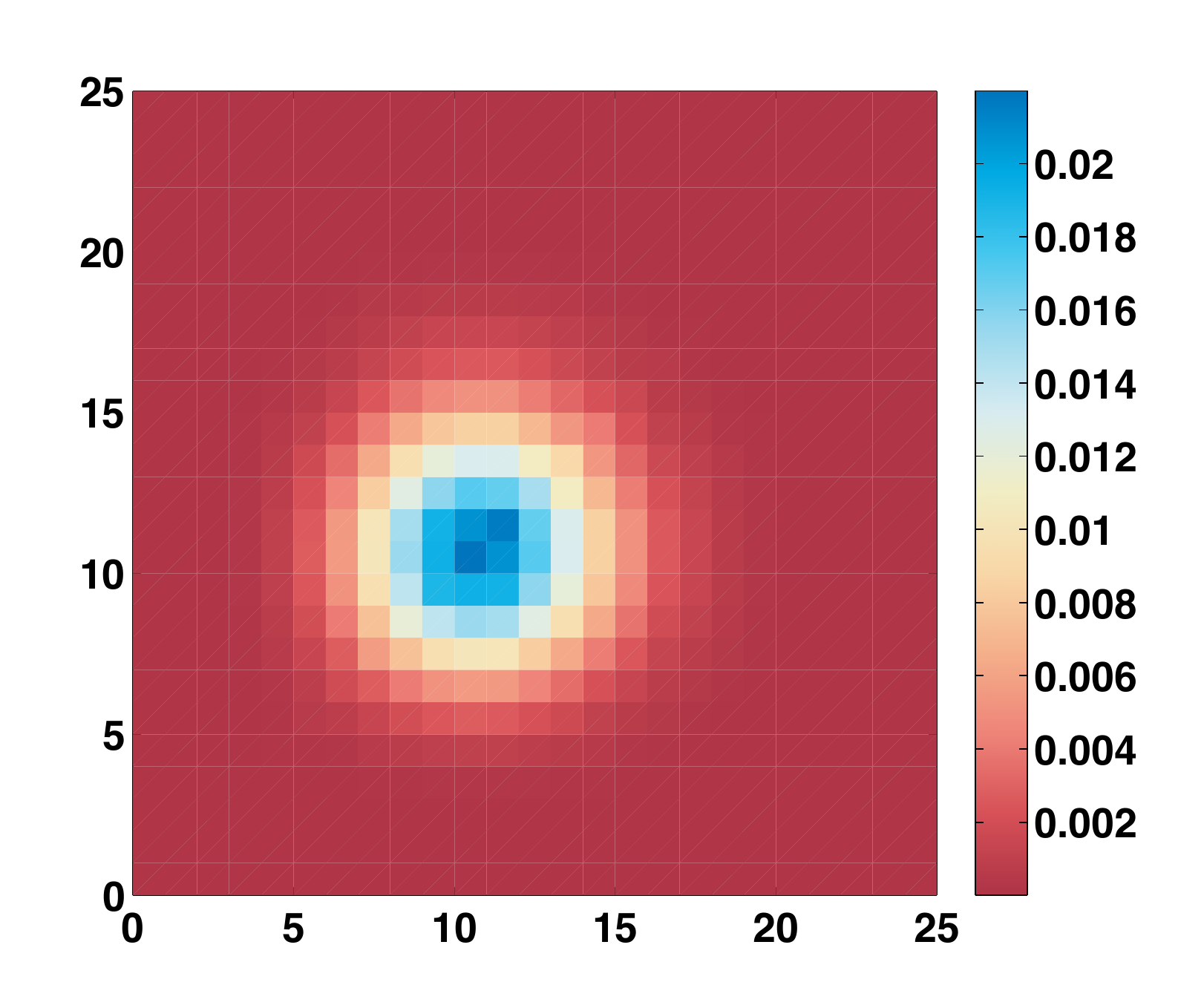}
  	\put(2,39.5){\rotatebox{90}{$k_2$}}
	\put(43,2){$k_1$}  
 \end{overpic}  
      \begin{overpic}[width=0.42\textwidth]{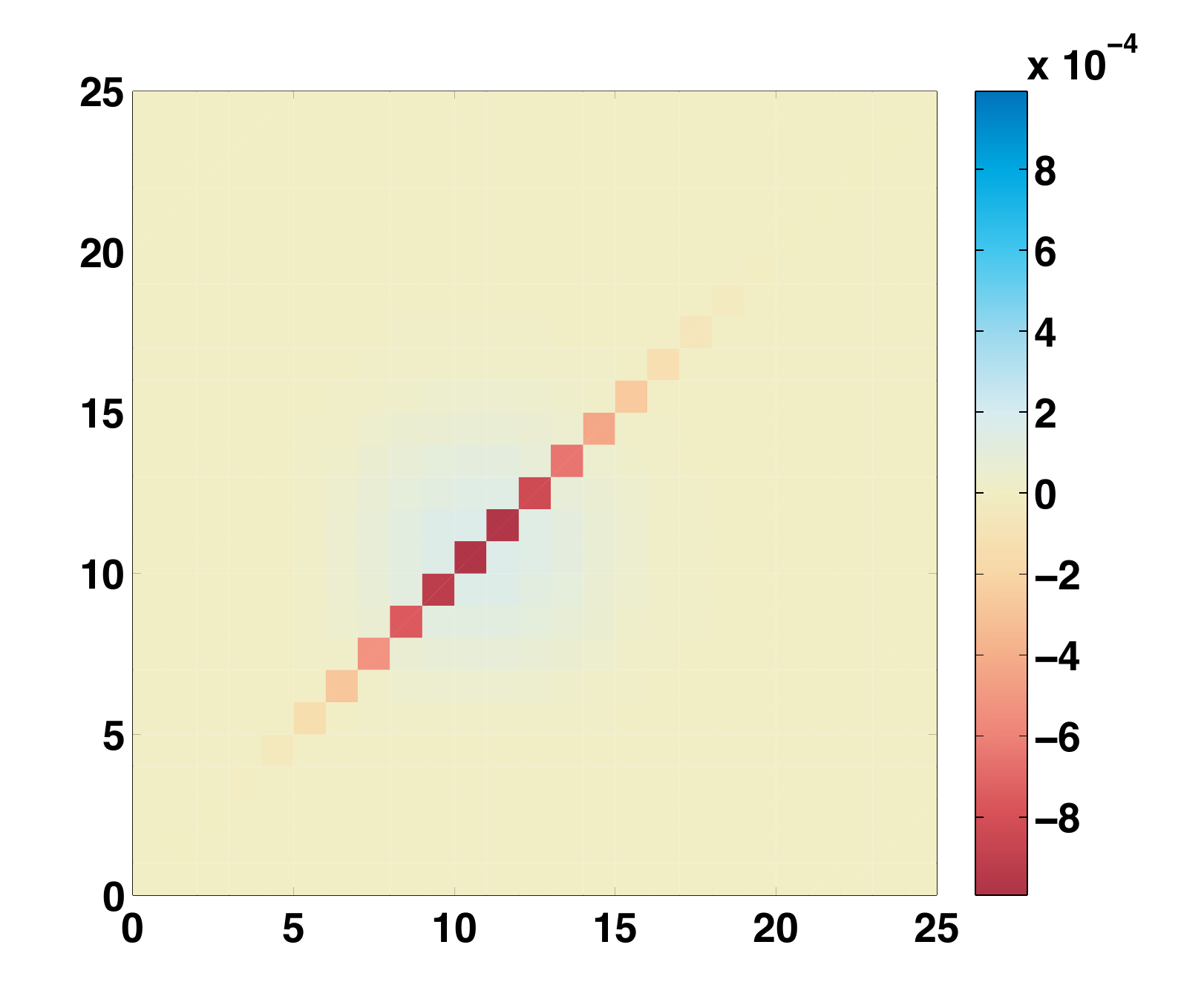}
  	\put(2,39.5){\rotatebox{90}{$k_2$}}
	\put(43,2){$k_1$}  
  \end{overpic}  
  \caption{\footnotesize{Numerical illustration of the
        validity of the mean-field assumption
        (\ref{eq_standard_molecular_chaos}). We average over 100
        realizations of Algorithm \ref{Algo_network_generation} using
        $n = 125$ particles and a time step of $\De t = 10^{-3}$.
        No new particles enter the
        system ($\rj = 0$), the edge creation rate is
        $\FCon(k_i,k_j)= 2$, and the edge deletion rate
        is $\FDel(k_i, k_j) =  k_i+k_j$.  We show
        (top) the 2-particle distribution $P_{k_1,k_2}^{(2)}(t)$,
        (middle) the product $P_{k_1}(t) P_{k_2}(t)$ of the 1-particle
        distributions, and (bottom) the difference
        $P_{k_1,k_2}^{(2)}(t) - P_{k_1}(t) P_{k_2}(t)$ at time
        $t=1/10$.         }
} 
  \label{fig_p_1_p_2_comparison}
\end{figure}

The 2-particle degree distribution $P^{(2)}_{k_1,k_2}$ is the probability that 
two nodes selected uniformly at random without replacement
have degrees $k_1$ and $k_2$. In our mean-field closure, we approximate this quantity by the product $P_{k_1}P_{k_2}$, where $P_{k_1}$ is the probability that a single node selected uniformly at random has degree $k_1$ (and $P_{k_2}$ is defined analogously). In Fig.~\ref{fig_p_1_p_2_comparison}, we compare the empirical distributions
$P_{k_1,k_2}^{(2)}(t)$ and $P_{k_1}(t)P_{k_2}(t)$ generated from
100 realizations of Algorithm \ref{Algo_network_generation} using 125
particles (and no node creation).

%%%%%%%%%%%%%%%   FIGURE    %%%%%%%%%%%%%%%
%%%%%%%%%%%%%%%%%%%%%%%%%%%%%%%%%%%%

We see that our mean-field approximation does well on this example, and the main error
occurs when $k_1=k_2$, for which the product $P_{k_1}(t)P_{k_2}(t)$ of 1-particle distributions is slightly larger than
the 2-particle distribution $P_{k_1,k_2}^{(2)}(t)$. This discrepancy arises because one cannot select the same
node twice when evaluating the correlation function
$P_{k_1,k_2}^{(2)}(t)$, so the probability of finding two nodes
with the same degree is lower than that estimated by
$P_{k_1}(t)P_{k_2}(t)$ (which corresponds to choosing two nodes
uniformly randomly with replacement). The difference should therefore tend to $0$ as $1/n$ as
the number $n$ of nodes becomes infinite.

One way to evaluate the difference between two probability
distributions is the Kolmogorov--Smirnov (KS) test
\cite{Peacock_1983}, which gives the probability $\KS$ that one rejects
the hypothesis that the two distributions are equal. 
For $P_{k_1,k_2}^{(2)}(t)$ and $P_{k_1}(t)P_{k_2}(t)$,
we find that $\KS \approx 2.0 \times10^{-3}$.

%%%%%%%%%%%%%%%%%%%%%%%%%%%%%%%%%%%%
\subsection{Example Scenario:  Local State Degree Distribution}\label{sec_mock_scenario}
%%%%%%%%%%%%%%%%%%%%%%%%%%%%%%%%%%%%

%%%%%%%%%%%%%%%%%%%%%%%%%%%%%%%%%%%%
\begin{figure}[h!]
%\iffalse
   \begin{overpic}[width=0.5\textwidth]{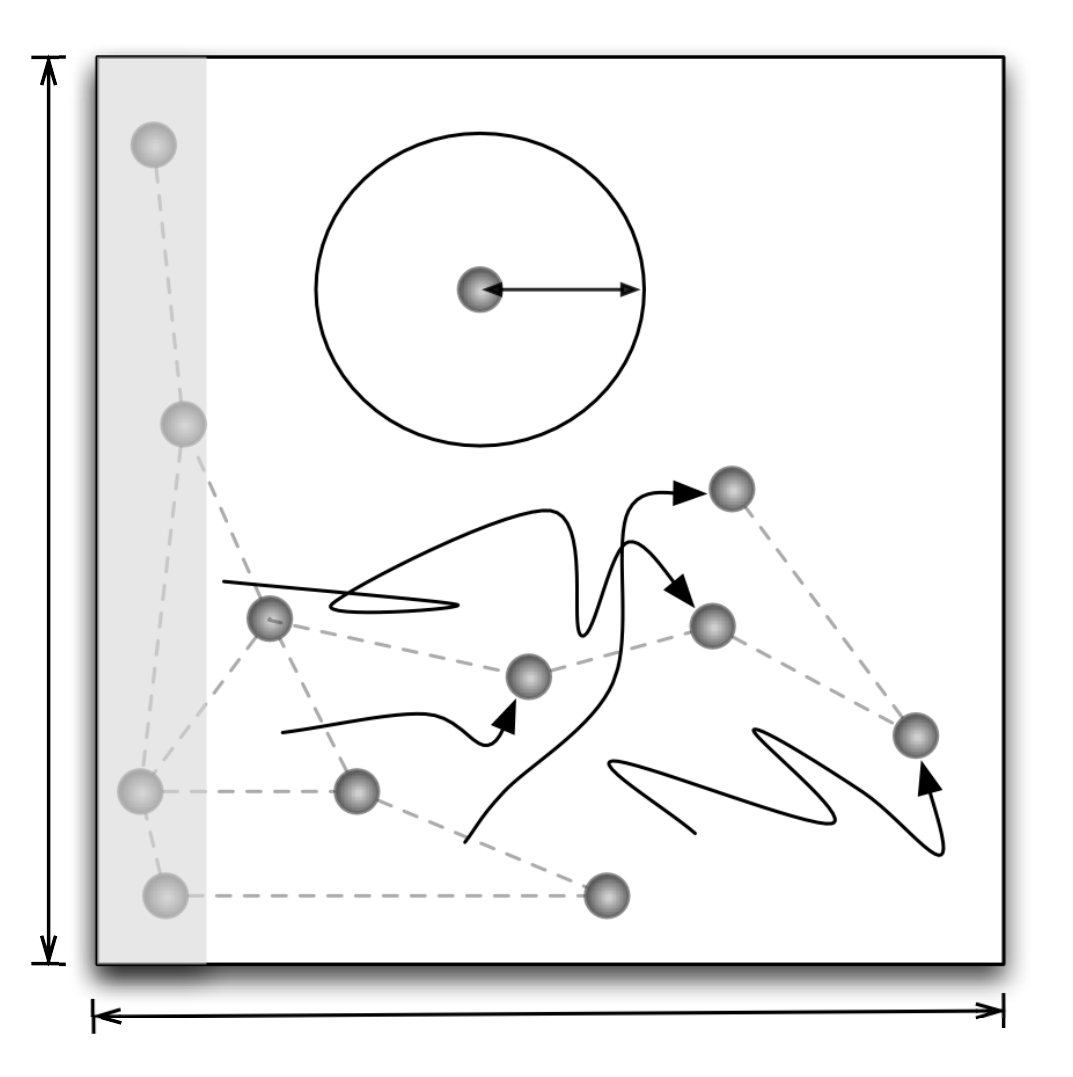} 
  	\put(50,1){\Large $x$}
	\put(-1,50){\rotatebox{90}{\Large $y$}}  
	\put(51.5,75){$\eps$}
	\put(55,59){Drift $\rightarrow$}
	\put(12,40){\rotatebox{90}{New node entry}}
  \end{overpic}  
%\fi
  \caption{\footnotesize{Illustration of the example scenario in Section \ref{sec_mock_scenario}. Nodes enter the system in a strip on the left-hand side of the unit square and then diffuse and drift to the right through equation \eqref{eq_drift_diffusion}. There are reflective boundary conditions at $x=0$ and $x=1$ and periodic boundary conditions at $y=0$ and $y=1$. We create edges between nearby nodes according to equation (\ref{eq_Con_choice_1}). Edges are deleted at the rate given by equation (\ref{eq_Del_choice}).}}
  \label{fig_mock_scenario_illustration}
\end{figure}
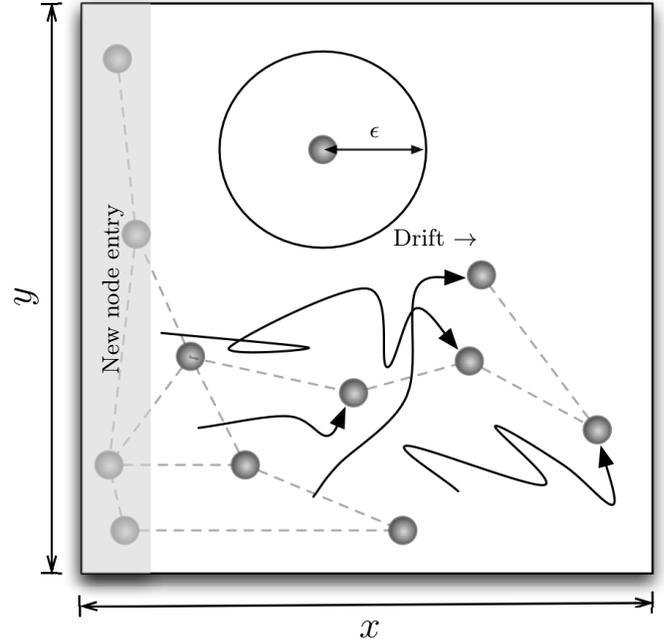
%%%%%%%%%%%%%%%%%%%%%%%%%%%%%%%%%%%%

The example in Section \ref{molchaos} was particularly simple, as it
focused only on the network aspect of the model.
We now want to compare Monte-Carlo simulations of a much more
complicated evolving spatial
network with a numerical solution of the reduced equation
\eqref{eq_del_addition}.
We select our new example to illustrate and evaluate all of the model components 
described in Table \ref{table_full_model}. It does not represent any particular physical or
biological process.  

Let's consider noninteracting point particles contained in the unit square, so
one can describe the state of each particle by its position vector
$\sv_i = (x_i,y_i)\in[0,1]^2$. 
We suppose that these positions evolve
according to the SDEs \cite{Oksendal_2014}  
\begin{equation}\label{eq_drift_diffusion}
	\dd X_i = \mu \,\dd t + \sigma\, \dd W_t \, , \quad \dd Y_i =
        \sigma\, \dd W_t , 
\end{equation}
where as before we use capital letters $\Sv_i = (X_i,Y_i)$ to distinguish
random variables from the values that they take. We assume that the
 drift coefficient $\mu>0$ and volatility
coefficient $\sigma>0$ are constant (corresponding to
a constant diffusion coefficient $\sigma^2/2$). To initialise, we place $1000$ particles with degree
$0$ uniformly at random in the rectangle
$(X_i,Y_i)\in[0,1/10]\times[0,1]$. We impose reflective boundary
conditions at  $x=0$ and $x=1$, and we impose periodic boundary conditions at $y=0$ and $y=1$.
To generate some spatial heterogeneity, we suppose that the rate of
edge creation between nodes depends both on the distance 
between nodes and on the spatial coordinates of each node. We take
\begin{equation}\label{eq_Con_choice_1}
	\FCon(\Sv_i, k_i , \Sv_j, k_j) =    \left\{ 
\begin{array}{cl} 
	X_i +X_j\,, & \text{  if   } ||\Sv_i - \Sv_j || \leq \eps \,, \\
	0\,, & \text{  otherwise} 
\end{array} \right\} .
\end{equation}
In contrast, we suppose that the rate of edge deletion 
per edge between nodes $i$ and $j$ depends on the degrees
of nodes $i$ and $j$ but is independent of position. Specifically, we take
\begin{equation}\label{eq_Del_choice}
	\FDel(\Sv_i, k_i , \Sv_j, k_j) = \frac{k_i+k_j}{10}.
\end{equation}
We introduce new nodes of degree $0$ at
a rate $\rj$ uniformly at random in the rectangle
$(x,y)\in[0,1/10]\times[0,1]$. Therefore,  
\begin{equation}\label{eq_newph_choice}
 	 \NewPh (x,y) = \left\{\begin{array}{ll}
 		 10 \,, & \mbox{ if }x<1/10\,,\\
 		 0 \,, & \mbox{otherwise\,.}\end{array}
 	 \right.
\end{equation}
We give a schematic illustration of these processes in
Fig.~\ref{fig_mock_scenario_illustration}.
We simulate the system using Algorithm \ref{Algo_network_generation} until final time
$T_{\text{end}}=1/2$.

Having defined the stochastic process that we are simulating, we now
turn to the reduced model (\ref{eq_del_addition}).
Because $\FCon(\sv_1, k_1 , \sv_2, k_2)$ and  $\FDel(\sv_1, k_1 , \sv_2,
k_2)$ are independent of $y_1$ and $y_2$, and $\NewPh(x_1,y_1)$ is
independent of  $y_1$, we expect a solution in which $u_{k_1}(t,\sv_1)$
is independent of $y_1$. Integrating over $y_2$ gives
\begin{widetext}
\iffalse
\begin{align}
\lefteqn{	 \frac{\p u_{k_1}}{\p t} (t, x_1) +  \mu \frac{\p
  u_{k_1}}{\p x_1} (t, x_1) 
  - \frac{\sigma^2}{2}\frac{\p^2 u_{k_1}}{\p x_1^2}(t, x_1)  = 
   \left(   \sum_{{k_2} = 0}^{\infty} \int_{0}^1   \hat{\FCon}
 (x_1, x_2 )    u_{k_2} (t,  x_2) \, \dd x_2
      \right)  \left(u_{k_1-1}  (t,  x_1) -u_{k_1} (t, x_1)\right)
 }\hspace{0.3cm} & \nonumber 
   \\ 
&\mbox{}+ \frac{1}{\int_{\mS}
  \sum_{k_1=0}^\infty k_1 
  u_{k_1}(t,\sv_1)\, \dd \sv_1} \sum_{ {k_2} = 1}^{\infty}
  \left( \int_{0}^1 u_{k_2} (t,  x_2) \, 
  \dd x_2 \right)k_2 \left(
  (k_1 + 1)\FDel ( k_1+1,  k_2)  u_{k_1+1}
  (t,x_1) 
-
  k_1\FDel ( k_1,
  k_2)  u_{k_1} (t,x_1)
\right)
 \\
&\mbox{ }
    + \rj\NewPh ( x_1) \de_{k_1, 0} .\label{eq_del_example}
\end{align}
\fi
\begin{align}
\lefteqn{	 \frac{\p u_{k_1}}{\p t} (t, x_1) +  \mu \frac{\p
  u_{k_1}}{\p x_1} (t, x_1) 
  - \frac{\sigma^2}{2}\frac{\p^2 u_{k_1}}{\p x_1^2}(t, x_1)  = 
   \left(   \sum_{{k_2} = 0}^{\infty} \int_{0}^1   \hat{\FCon}
 (x_1, x_2 )    u_{k_2} (t,  x_2) \, \dd x_2
      \right)  \left(u_{k_1-1}  (t,  x_1) -u_{k_1} (t, x_1)\right)
 }\hspace{0.3cm} & \nonumber 
   \\ 
&\mbox{ }+\frac{(k_1+1)^2}{10} u_{k_1+1}(t,x_1) - \frac{k_1^2}{10} u_{k_1}(t,x_1) 
 + 
\frac{\int_{\mS} \sum_{k_2=1}^\infty k_2^2 
  u_{k_2}(t,\sv_2)\, \dd \sv_2}{\int_{\mS} \sum_{k_2=1}^\infty k_2 
  u_{k_2}(t,\sv_2)\, \dd \sv_2}\left(
  \frac{k_1 + 1}{10} u_{k_1+1}(t,x_1) -
  \frac{k_1}{10}  u_{k_1} (t,x_1)
\right)
\nonumber  \\
&\mbox{ }
    + \rj\NewPh ( x_1) \de_{k_1, 0},\label{eq_del_example}
\end{align}
where
\begin{equation}\label{eq_Con_choice_2} 
\hat{\FCon}(x_1,x_2) =    \left\{ 
\begin{array}{cl} 
	2(x_1+x_2) \sqrt{\eps^2 -  |x_1 - x_2 |^2}\,, & \text{  if   }
                                                        |x_1 - x_2 |
                                                        \leq \eps\,,
  \\ 
	0\,, & \text{  otherwise} 
\end{array} \right\}\,.
\end{equation}
\end{widetext}
We solve (\ref{eq_del_example}) with no-flux conditions at $x=0$ and $x=1$ and initial
condition
\begin{equation}\label{eq_mock_scen_IC}
	u_{k_1}(t=0,x_1) = \left\{ \begin{array}{ll} 10^4 \de_{k,0}\,,  & x_1 \in [0,1/10]\,, \\
	0\, & \mbox{otherwise}\,,\end{array}
\right.
\end{equation}
because initially there are $10^3$ particles placed uniformly at random in
$[0,1/10]\times[0,1]$.

\begin{figure}[h!]
   \begin{overpic}[width=0.45\textwidth]{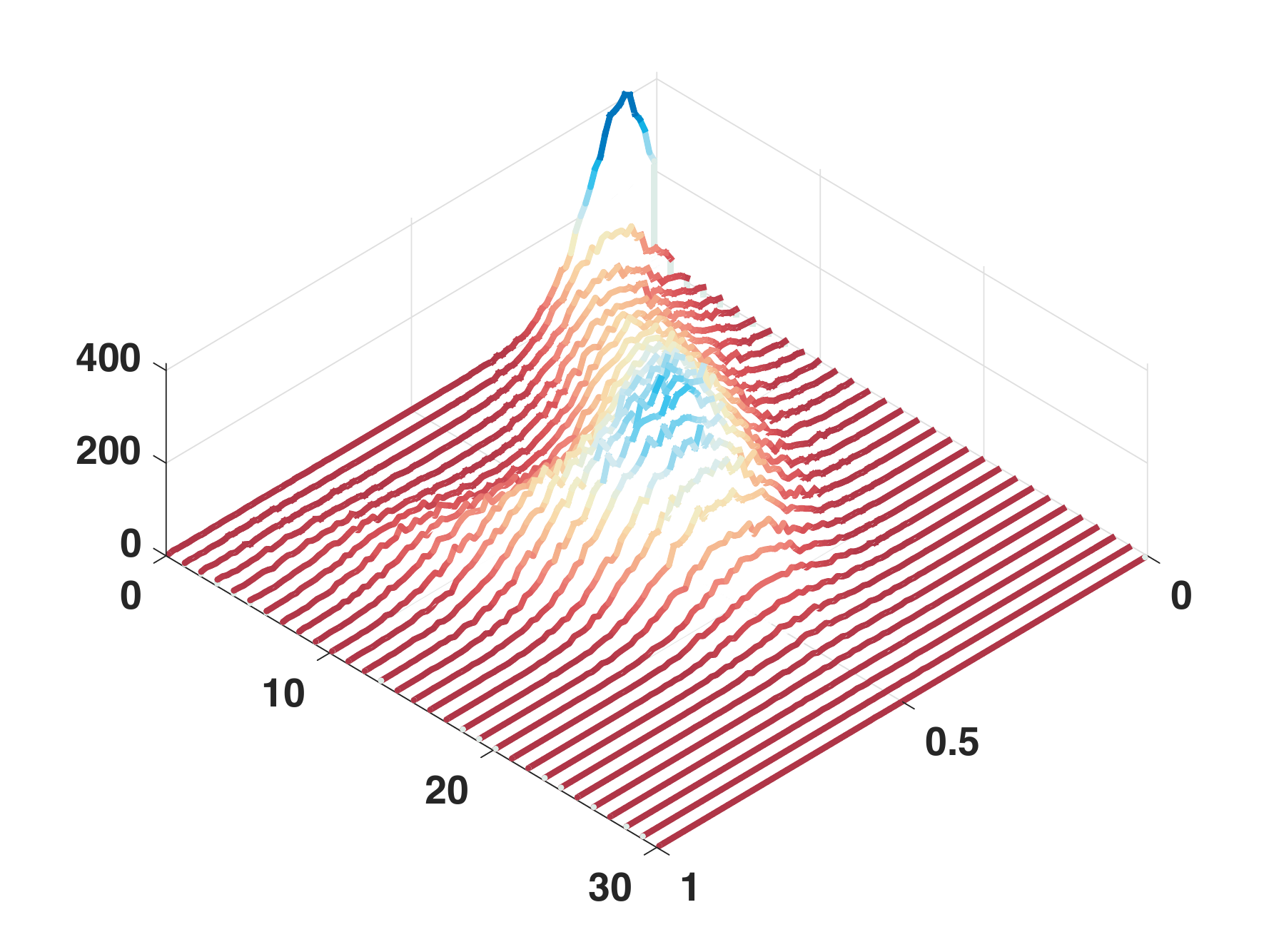} 
  	\put(71,8){\rotatebox{34}{\bf{$x$ position}}}
	\put(17,16){\rotatebox{-34}{\bf{degree}}}  
  \end{overpic}
    \begin{overpic}[width=0.45\textwidth]{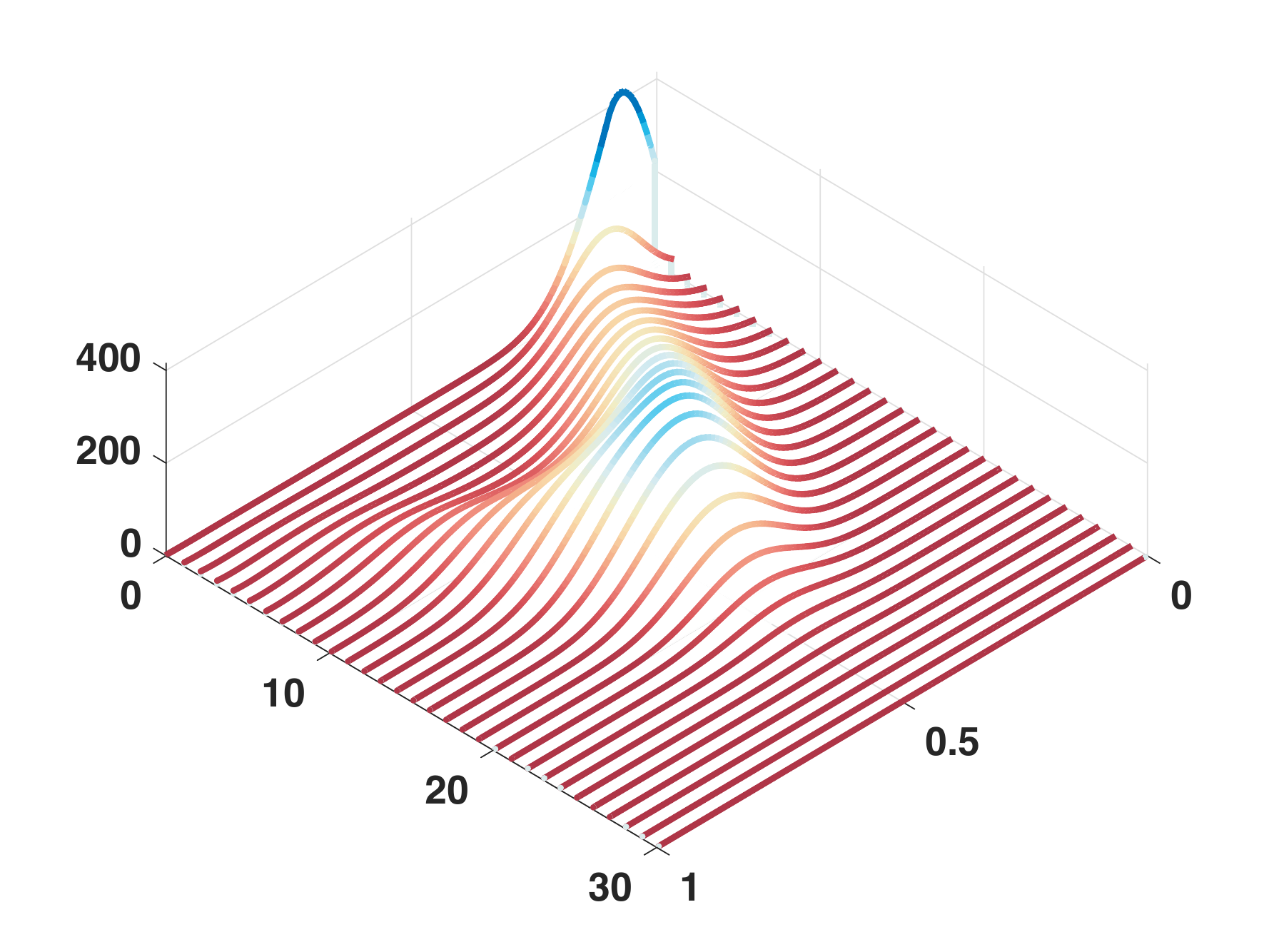}
  	\put(71,8){\rotatebox{34}{\bf{$x_1$}}}
	\put(17,16){\rotatebox{-34}{\bf{$k_1$}}}  
  \end{overpic}  
  \caption{\footnotesize{Comparison of (top) the mean of 200
        Monte-Carlo simulations using Algorithm
        \ref{Algo_network_generation} binned into compartments of size
        $1/100$ along the $x$-axis and (bottom) $u_{k_1}(1/2,x_1)$
        obtained from the numerical solution
        of equation \eqref{eq_del_example}. The parameter values are 
 $\rj = 500$, $\mu = 3/4$, $\sigma = 1/4$, and $\eps = 0.1$; and the
 time step in the stochastic (i.e., Monte-Carlo) simulations is $\De t =
      10^{-4}$. We initialize the simulations 
      with $10^3$ particles placed uniformly randomly in the domain
      $[0,1/10]\times[0,1]$. We show results at the final time 
 $T_{\text{end}} = 1/2$.}
 } 
  \label{fig_mock_scenario}
\end{figure}

In Fig.~\ref{fig_mock_scenario}, we show a comparison between a Monte-Carlo simulation of $u_k(t,x)$ using Algorithm
\ref{Algo_network_generation} and a numerical solution of equation
(\ref{eq_del_example}) using a second-order central difference
finite-volume method (FVM) in space and a fourth-order Runge--Kutta
scheme in time. We use the parameter values $\mu =
3/4$, $\sigma = 1/4$, $\rj = 500$, and $\epsilon = 0.1$.

We observe that the distribution function $u_k(t,x_1)$ is nontrivial, and
that the reduced model does a good job of capturing the empirical distribution
that we obtain from Monte-Carlo simulations. The KS probability for the
distributions in Fig.~\ref{fig_mock_scenario} is $\KS \approx
3.2\times10^{-2}$.

%%%%%%%%%%%%%%%%%%%%%%%%%%%%%%%%%%%%
\subsection{Limit of Large Networks}\label{sec_Boltzmann_param}
%%%%%%%%%%%%%%%%%%%%%%%%%%%%%%%%%%%%

We expect our mean-field assumption to be more accurate for
larger networks. We now briefly investigate this hypothesis in the context of degree distributions. (Naturally, it is also relevant to consider this hypothesis for other quantities.)

For a network with $n$ nodes, $u_{k} = O(n)$, so we see that
equation (\ref{eq_del_addition}) converges to a sensible limit as
$n \rightarrow \infty$ if $\FCon =O(1/n)$ and $\FDel=O(1)$.
To investigate the convergence for large $n$, we consider
networks in which the number of nodes is constant in time (i.e., $\rj\equiv
0$). We first consider networks in which there is no edge deletion
(i.e., $\FDel\equiv 0$). We take the edge-creation rate to be $500/n$ times
that given in equation (\ref{eq_Con_choice_1}). We choose all other parameters as in Section \ref{sec_mock_scenario}.

\begin{figure}[h!]
   \begin{overpic}[width=0.49\textwidth]{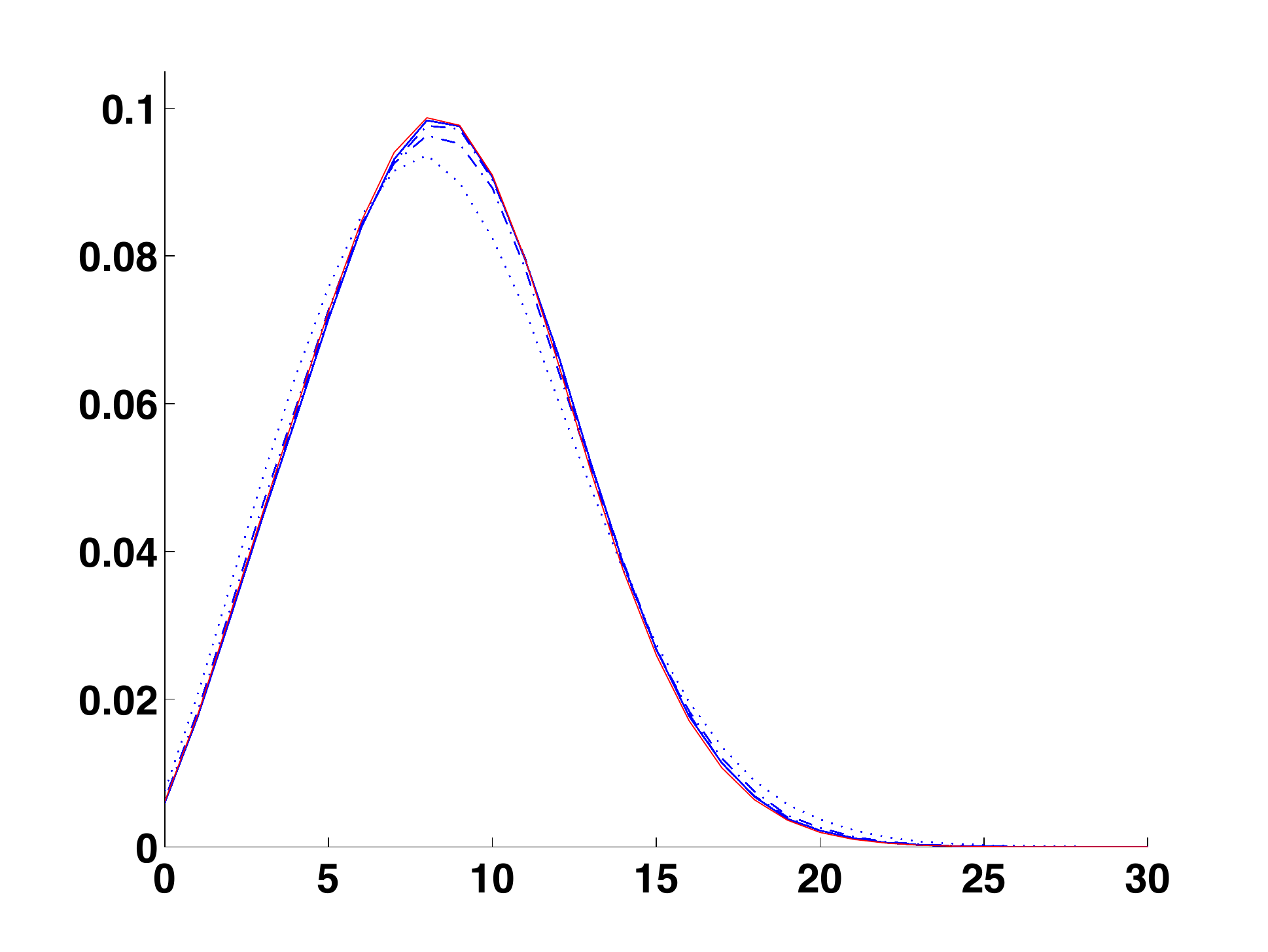} 
  	\put(0,27){\rotatebox{90}{probability density}}
	\put(45,0){degree} 
  \end{overpic} \\ \hspace{5pt} 
    \begin{overpic}[width=0.49\textwidth]{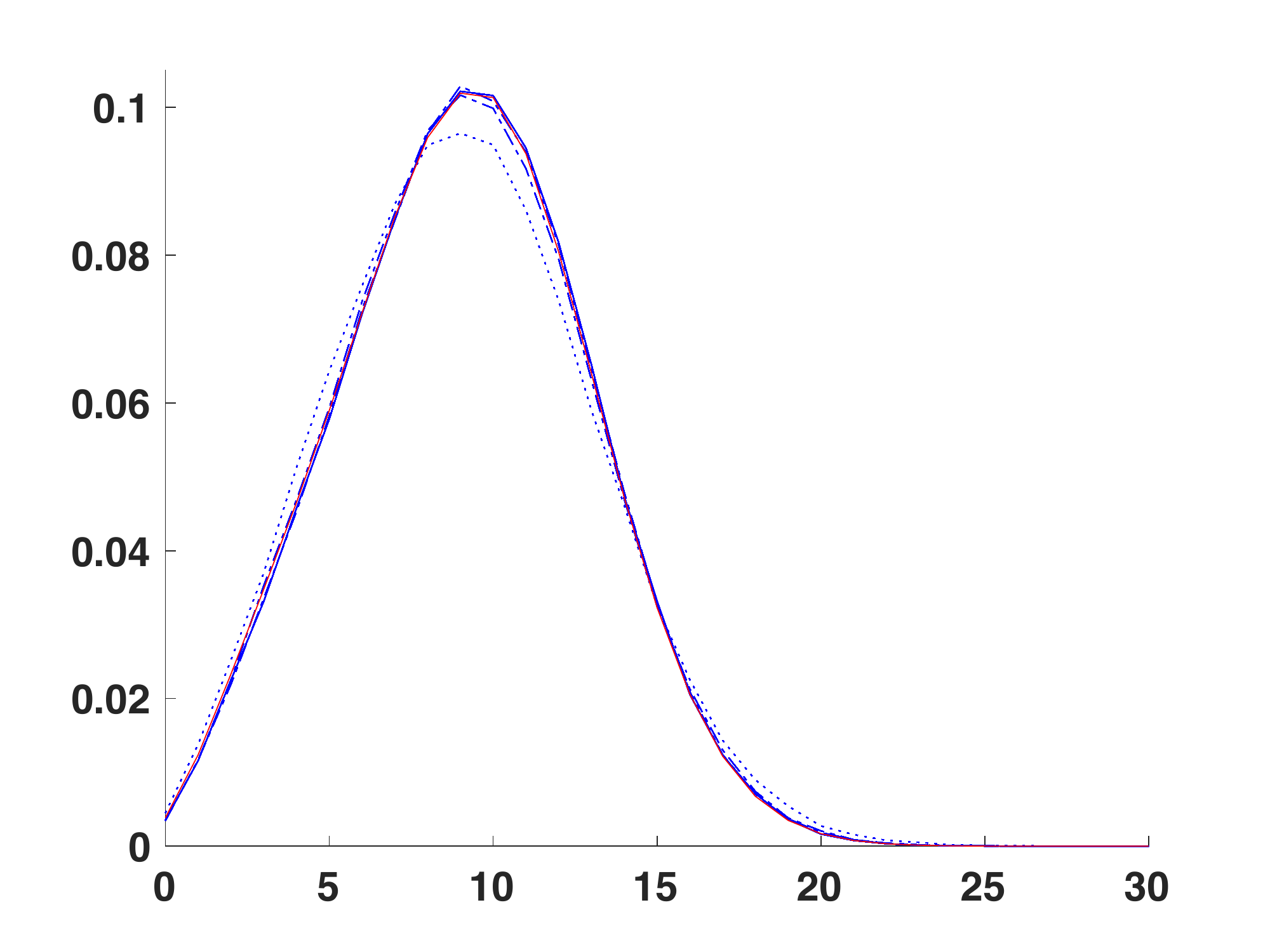}
  	\put(0,27){\rotatebox{90}{probability density}}
	\put(45,0){degree}  
  \end{overpic}  
  \caption{\footnotesize{Comparison of degree
        distribution at time $T_{\text{end}}=1/2$ determined from
        equation (\ref{eq_del_example}) (red) 
        and from a mean over multiple Monte-Carlo simulations
        using Algorithm \ref{Algo_network_generation} with a time step
        of $\De t = 10^{-3}$ (blue). 
With $n$ nodes (constant in time; $\rj\equiv 0$), we average the simulations are over $10^6/n$ realizations, where $n=125$
  (dotted curve), $n=500$ (dash-dotted 
  curve), $n=2000$ (dash-dotted curve), and $n=8000$ (solid
  curve). 
(Top) No edge deletion (i.e., $\FDel \equiv 0$), and the edge-creation rate
is $500/n$ times that given in  
  equation \eqref{eq_Con_choice_1}.
(Bottom) The edge deletion is given by equation (\ref{eq_Del_choice}), and the edge creation rate
$750/n$ times that given in  
  equation \eqref{eq_Con_choice_1}.
The other parameter values are as in Section \ref{sec_mock_scenario}.
    } } 
  \label{fig_Boltzmann_const_convergence}
\end{figure}

In the top panel of Fig.~\ref{fig_Boltzmann_const_convergence}, we show the mean 
  degree distribution sampled over $10^6/n$ realizations of Algorithm
  \ref{Algo_network_generation} for $n = 125$, $n = 500$, $n = 2000$, and
  $n = 8000$. We also show the degree distribution calculated by solving the IPDE
  (\ref{eq_del_example}). Qualitatively, this figure supports the
  hypothesis that, at least in terms of degree distribution, the
  mean-field approximation is more accurate for larger networks.

We now introduce edge deletion and choose $\FDel$ to be given by
equation (\ref{eq_Del_choice}). We increase the rate of edge creation slightly
by taking the edge-creation rate to be $750/n$ times
that given in equation (\ref{eq_Con_choice_1}). We show the resulting mean 
  degree distribution sampled over $10^6/n$ realizations of Algorithm
  \ref{Algo_network_generation} for $n = 125$, $n = 500$, $n = 2000$, and
  $n = 8000$ in the bottom panel of Fig.~\ref{fig_Boltzmann_const_convergence}(bottom) along with the degree distribution that we calculated by solving the IPDE (\ref{eq_del_example}).

%%%%%%%%%%%% the next figure is commented-out %%%%%%%%

\iffalse
\begin{figure} 
\centering 
\begin{overpic}[width=0.4\textwidth]{N_vs_rho_KS.pdf}
  	\put(-3,43){\rotatebox{90}{$\KS$ value}}
	\put(30,2){Number of nodes, $n$}  
  \end{overpic}
  \caption{\footnotesize{For the degree distribution curves given in Fig.~\ref{fig_Boltzmann_const_convergence}, we plot values of $\KS$ calculated from Monte Carlo simulations for varying $n$ when compared to the degree distribution calculated from equation (\ref{eq_del_addition}). In the solid blue curve, we include only edge creation; in the dashed red curve, we include both edge creation and edge deletion.}}
  \label{fig_KS_decay}
\end{figure}
\fi

%%%%%%%%

%%%%%%%%%%%%%%%%%%%%%%%%%%%%%%%%%%%%
\subsection{One-Step Network Creation Versus Kinetic Approximations}\label{sec_one_step_kinetic}
%%%%%%%%%%%%%%%%%%%%%%%%%%%%%%%%%%%%

Dynamic models of network creation can provide an alternative to
one-step network creation.
 For example, in the standard $G(n,p)$ Erd\H{o}s--R\'{e}nyi (ER)
model \cite{Newman_2010}, one specifies that a network has $n$ nodes and that
each pair of nodes is connected with independent, constant probability
$p\in(0,1)$. This leads to a binomial degree distribution
$\text{Bin}(n-1,p)$, which becomes the Poisson distribution
 $\text{Pois}(np)$ in the limit $n\to\infty$ with fixed $np$.
 Reference~\cite{Krapivsky_2010} discussed an alternative,
dynamic approach to the ER model in which an initially unconnected
network has $n$ nodes and each node connects to other nodes
 uniformly at random at a specified rate. In the $n \to \infty$ limit, one can solve a
master equation to obtain a Poisson degree
distribution, which coincides with the standard model when halted at a
specific time (depending on the edge-creation rate).
We note also the work by Krioukov and Ostilli
\cite{Krioukov_2013}, who showed that certain equilibrium
ensembles create the same distribution of graphs as nonequilibrium
ensembles.

References~\cite{Boguna_2003, Boguna_2003_b} discussed a one-step network-creation model
that allows nodes to have an associated state.  Nodes $i$ and $j$ have randomly distributed latent social variables $\vect{s}_i$ and $\vect{s}_j$, and these nodes are adjacent to each other with probability
$r(\vect{s}_i,\vect{s}_j)$.  References~\cite{Boguna_2003, Boguna_2003_b} then used a mean-field approximation to derive a formula for the degree distribution (depending on the probability distribution of the social variables and on the function $r$) of the network as the number of nodes tends to infinity.

\begin{figure}[h!]
   \begin{overpic}[width=0.49\textwidth]{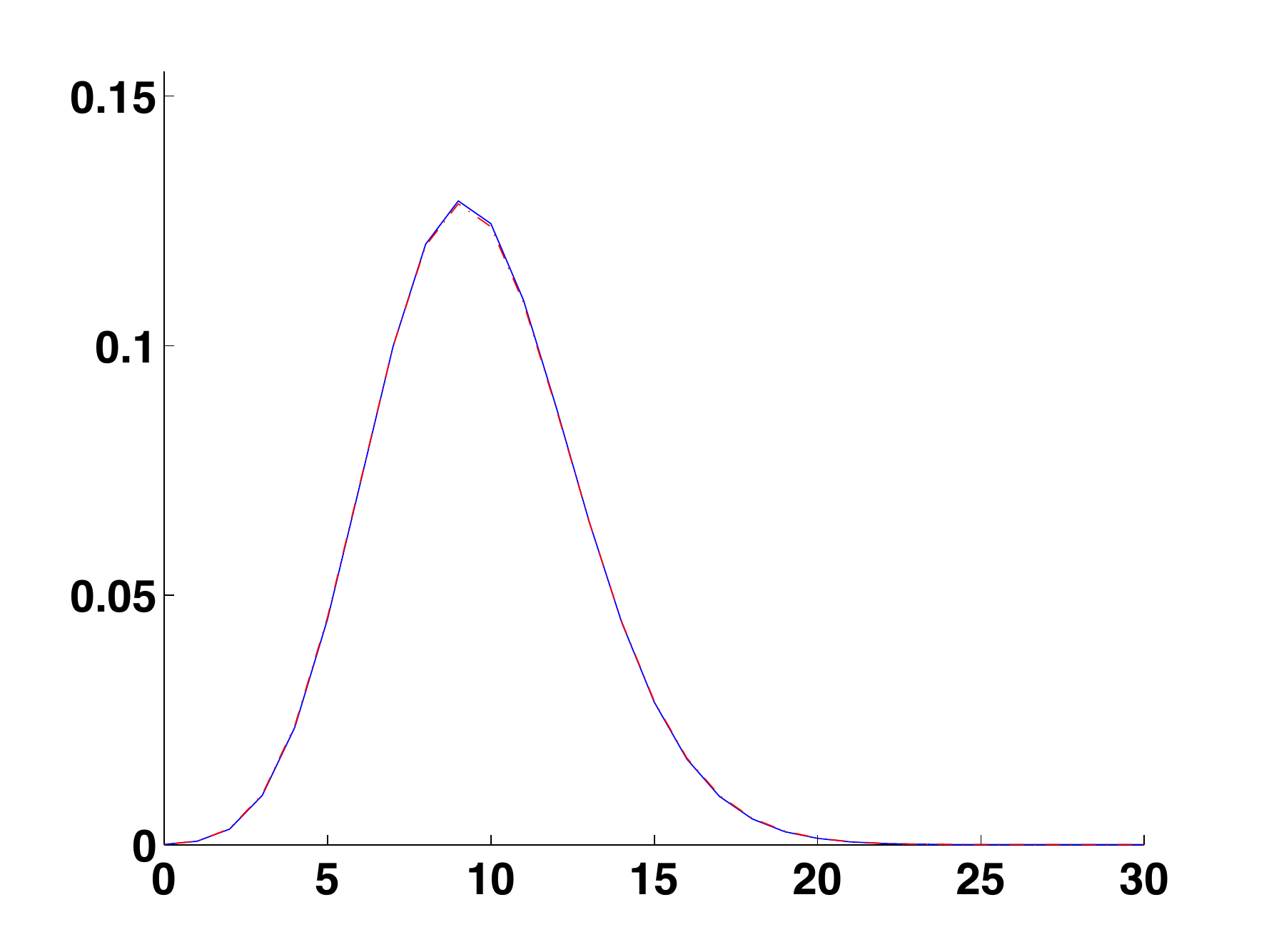} 
  	\put(0,24){\rotatebox{90}{probability density}}
	\put(45,0){degree}  
  \end{overpic} 
  \caption{\footnotesize{Analytical approximation to the degree
      distribution (solid blue curve) of the latent social-space
      model given in Ref.~\cite{Boguna_2003} versus the result from a kinetic
      formulation (red dash-dotted curve) using equation
      (\ref{eq_final_IPDE}). We uniformly distribute nodes with a
      social parameter $h\in[0,125]$. We use the parameter values
      $n=1000$, $\alpha = 3$, and $b = 1/2$. The probability of
      connection and choice of $\FCon$ is given by equation
      \eqref{eq_boguna_conn}. There is no edge deletion (i.e., $\FDel \equiv
      0$), and the system is of constant size (i.e., $\rj \equiv 0$). We solve the
      kinetic equation until final time $T_{\text{end}} =
      1$.}}  
  \label{fig_boguna_comparison}
\end{figure}

One can use our kinetic approach as an alternative to
one-step creation to examine such a scenario. To demonstrate this, we consider
the case investigated in \cite{Boguna_2003}, where the latent
social variable is a positive 
scalar \mbox{$\vect{s} = h\in [0,h_{\max}]$} and  
\begin{equation}\label{eq_boguna_conn}
	r(h_i,h_j) = \frac{1}{1 + (b^{-1}|h_i - h_j |)^{\alpha}}
\end{equation}
for constants $b$ and $\alpha$. We solve equation \eqref{eq_final_IPDE} with an initial condition
corresponding to $n$ unconnected nodes that are distributed
uniformly at random in a state space $[0,h_{\max}]$.
No new nodes enter the system (i.e., $\rj\equiv0$), there is
no edge deletion (i.e., $\FDel \equiv 0$), 
and the edge-creation rate is $\FCon = r(h_i,h_j)$.  Thus, at time 
$T_{\text{end}}=1$, the expected number of edges between node $i$ and node $j$ is
$r(h_i,h_j)$. In Fig.~\ref{fig_boguna_comparison}, we see for parameter values
$n=1000$, $\alpha = 3$, and $b = 1/2$ that the degree distribution
given by our mean-field model at $t = T_{\text{end}}$ matches very closely with
the analytical formula of \cite{Boguna_2003}. We solved the IPDE using a method of lines with a spatial discretization of $\De h = 1/4$ and a fourth-order Runge--Kutta scheme in time.

%%%%%%%%%%%%%%%%%%%%%%%%%%%%%%%%%%%%
\subsection{Final Remarks: Moment
  Closure}\label{sec_numerics_finishing_remarks} 
%%%%%%%%%%%%%%%%%%%%%%%%%%%%%%%%%%%%

Despite the fact that equations \eqref{eq_final_IPDE} and \eqref{eq_del_addition} are of much lower
dimension than equations \eqref{eq_HFP_l1} and (\ref{eq_full_HFP}), each of the former equations is still an
infinite system of IPDEs, and they may still be too
expensive to solve numerically. One can make further approximations by considering moments of the
density $u_k(t,\sv)$ with respect to the degree $k$:
\begin{equation}\label{eq_mon_moments}
	\mathcal{M}^{(r)}(t,\sv) = \sum_{k = 0}^\infty k^r u_k (t,\sv) \, .
\end{equation}
In particular, $\mathcal{M}^{(0)}$ gives the number density of nodes
(for which we obtain a closed equation if $\rj\equiv 0$ and
$\LO^{(1)}$ does not depend on the network structure), and 
$\mathcal{M}^{(1)}$ gives the number density multiplied by the mean
degree of a node with state vector $\sv$.  
It is only possible to obtain a closed system of equations for particular choices of $\FCon$ and
$\FDel$. In general, one needs to truncate the hierarchy of moment equations and then apply another closure assumption by positing an expression for a high-order moment in terms of lower-order moments \cite{Hillen_2003, Hillen_2004, Kuehn_2015}.

%%%%%%%%%%%%%%%%%%%%%%%%%%%%%%%%%%%%%%%%%%%%
%%%%%%%%%%%%%%%%%%%%%%%%%%%%%%%%%%%%%%%%%%%%
%%%%%%%%%%%%%%%%%%%%%%%%%%%%%%%%%%%%%%%%%%%%
%%%%%%%%%%%%%%%%%%%%%%%%%%%%%%%%%%%%%%%%%%%%
%%%%%%%%%%%%%%%%%%%%%%%%%%%%%%%%%%%%%%%%%%%%
\section{A Model for Osteocyte Network Formation}\label{sec_first_model_osteo}
%%%%%%%%%%%%%%%%%%%%%%%%%%%%%%%%%%%%%%%%%%%%
%%%%%%%%%%%%%%%%%%%%%%%%%%%%%%%%%%%%%%%%%%%%
%%%%%%%%%%%%%%%%%%%%%%%%%%%%%%%%%%%%%%%%%%%%
%%%%%%%%%%%%%%%%%%%%%%%%%%%%%%%%%%%%%%%%%%%%
%%%%%%%%%%%%%%%%%%%%%%%%%%%%%%%%%%%%%%%%%%%%
%%%%%%%%%%%%%%%%%%%%%%%%%%%%%%%%%%%%%%%%%%%%
%%%%%%%%%%%%%%%%%%%%%%%%%%%%%%%%%%%%%%%%%%%%
%%%%%%%%%%%%%%%%%%%%%%%%%%%%%%%%%%%%%%%%%%%%
%%%%%%%%%%%%%%%%%%%%%%%%%%%%%%%%%%%%%%%%%%%%
%%%%%%%%%%%%%%%%%%%%%%%%%%%%%%%%%%%%%%%%%%%%

\subsection{Osteocytes}\label{section_osteocyte_introduction}
%%%%%%%%%%%%%%%%%%%%%%%%%%%%%%%%%%%%%%%%%%%%

An \emph{osteocyte} \cite{KnotheTate_2004} is a dendritic cell in both cortical bone (dense, weight-bearing bone) and trabecular bone (flexible, highly vascular bone). The protrusions (i.e., dendrites) of the cell
are known as \emph{processes}, and they form a communication network between
the osteocytes and cells on the bone surface. To avoid confusion with the word ``processes'', which is common jargon in physics, we
henceforth refer to \emph{processes} as ``dendrites''. Osteocytes are densely packed
in bone. They occupy spherical spaces called lacunae, and their
dendrites occupy tunnels called canaliculi \cite{Carter_2014,
  Piattelli_2014}. The dendrites enable communication via gap
junctions, across which signaling molecules can diffuse. The precise purpose of this communication network is not known, but there are many conjectures. It has been suggested that the exchange of signaling molecules relates to
skeletal unloading, fatigue damage, and estrogen deficiency
\cite{Jilka_2013}. The range of signaling molecules that have been
detected is vast, and many also arise in the regulation of other
organs. These include Receptor Activator of Nuclear Factor Kappa-B
Ligand (RANKL) \cite{Jilka_2013}, Vascular Endothelial Growth Factor (VEGF)
\cite{Jilka_2013}, Parathyroid Hormone (PTH) \cite{Xiong_2014},
calcium ions ($\text{Ca}^{2+}$) \cite{Ishihara_2012}, and Sclerostin
\cite{Sapir_Koren_2014}. Additionally, there is a thin layer of fluid around the osteocyte network. Perturbations to osteocyte-network
organization can impact both fluid flow and diffusion, and they thereby
allow mechanosensation and signaling \cite{Kerschnitzki_2013}. 

\begin{figure}[h!]
%\iffalse
   \begin{overpic}[width=0.45\textwidth]{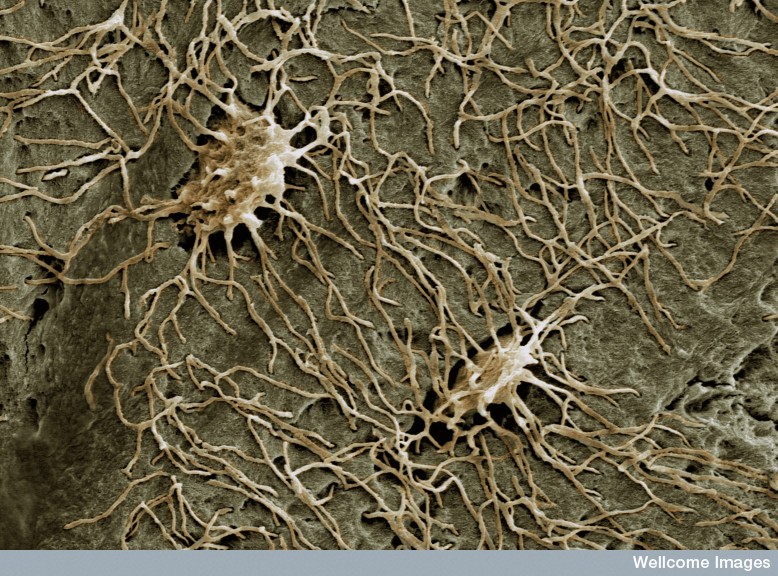} 
   \put(38.5,93){\scriptsize \textcolor{white}{A}} 
  \end{overpic}
%\fi
\caption{\footnotesize{Scanning electron micrograph of osteocytes in
    bone. The sample was prepared by embedding the bone in resin,
    which was subsequently etched with perchloric acid. The image was
    created by removing the entire mineral in the sample, leaving a
    replica of the cells. Therefore, what is observed is the resin
    that filled the spaces in the bone and the spaces inside the
    cells. (This picture is copyrighted work and is available via
    Creative Commons \cite{creativecommons} from Kevin Mackenzie,
    University of Aberdeen, 
    Wellcome Images \cite{Wellcome_Osteocyte_2010}.)}
}
  \label{fig_licence_free_scan}
\end{figure}

Because of the location of osteocyte networks in bone, it is difficult
to examine them experimentally. In Fig.~\ref{fig_licence_free_scan},
we show the outcome of applying an obtrusive experimental technique to
view osteocytes after nearby mineral has been
dissolved. Three-dimensional imaging data is now available by using
confocal microscopy \cite{Kamioka_2001, Sugawara_2011, Sugawara_2005},
and such work has led to the identification of some structural
features of osteocyte networks. Identified features include the mean
number of dendrites that protrude from each osteocyte
\cite{Sugawara_2005} and mean lengths of a canicular network
\cite{Buenzli_2015_b}. 
Other work has reported that high-density networks correlate
positively with high bone quality. Note that the mineral matrix has an
orientation (from dendrites and collagen), and bone quality is
associated with the level of organization of this mineral matrix
\cite{Kerschnitzki_2013}.

%%%%%%%%%%%%%%%%%%%%%%%%%%%%%%%%%%%%
%%%%%%%%%%%%%%%%%%%%%%%%%%%%%%%%%%%%

%%%%%%%%%%%%%%%%%%%%%%%%%%%%%%%%%%%%%%%%%%%%
\subsection{Formation Process}\label{section_formation_process}
%%%%%%%%%%%%%%%%%%%%%%%%%%%%%%%%%%%%%%%%%%%%

On the bone-tissue interface, two cell types are actively involved in the bone-formation process: \emph{osteoblasts} and \emph{osteoclasts}. Osteoblasts form a layer on the bone surface and secrete the osteoid bone matrix. The larger multi-nucleated osteoclasts subsequently resorb the bone matrix \cite{Robling_2006}.
Osteoblasts also express RANKL and osteoprotegerin (OPG), which promote and inhibit the bone resorption by osteoclasts, respectively. This is one example for how osteoblasts tightly regulate bone formation and destruction. 
 As the osteoblasts produce the calcium matrix, occasionally they
  become embedded within the bone. These osteoblasts then change
  morphology to become star-shaped osteocytes.  
  
Osteoblasts originate from mesenchymal cells and have one of four possible fates: undergo apoptosis (approximately 65\%), become embedded in bone as osteocytes (approximately 30\%), transform into inactive osteoblasts and become bone-lining cells, or transdifferentiate into cells that deposit chondroid bone \cite{Franz-Odendaal_2006}. Upon some signaling event, osteoclasts arrive at a bone and the osteoblasts move aside. The osteoclasts then burrow into the bone; as they do so, they resorb some of the osteocyte matrix. It has been suggested that after an osteocyte undergoes apoptosis, pro-osteoclastogenic signals are released by the osteocyte's neighbors in the network \cite{Kennedy_2014}. A trail of osteoblasts then follows the osteoclasts and secretes new bone matrix \footnote{The unit that consists of osteoblasts following osteoclasts is known as a ``Bone Multicellular Unit'' (BMU).}, although some of these get left behind to become osteocytes. Kamioka et al. suggested that osteoblasts are incorporated into a network by osteocytes extending their dendrites towards the osteoblast layer \cite{Kamioka_2001}.

\begin{figure}[h!]
   \begin{overpic}[width=0.38\textwidth]{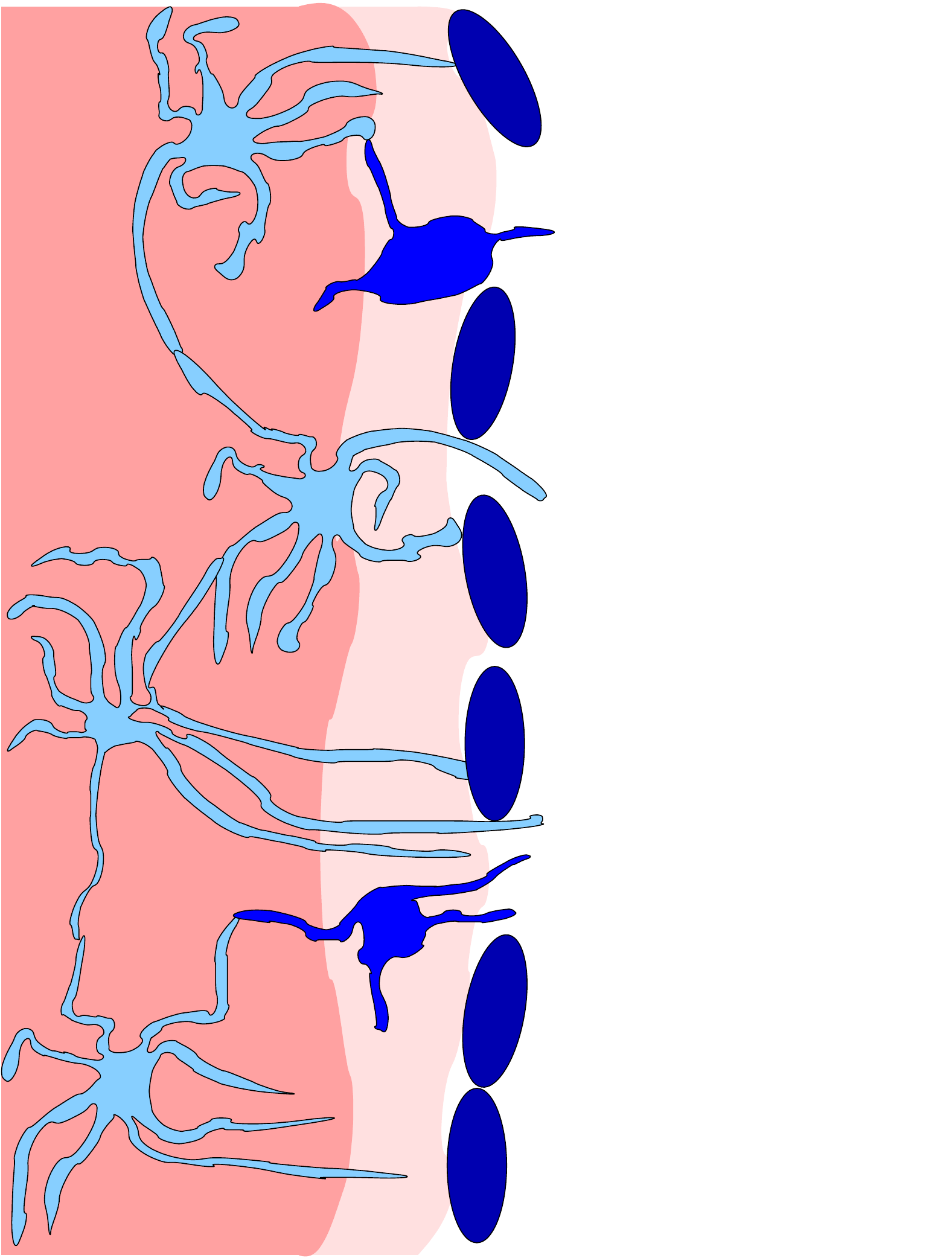} 
   	\put(38.5,93){\scriptsize \textcolor{white}{A}} 
	\put(38,40.5){\scriptsize \textcolor{white}{A}} 
	\put(37.5,70){\scriptsize \textcolor{white}{B}} 
	\put(38.5,54){\scriptsize \textcolor{white}{B}} 
	\put(38.5,18.5){\scriptsize \textcolor{white}{B}} 
	\put(37,7){\scriptsize \textcolor{white}{B}} 
  \end{overpic}
   \begin{overpic}[width=0.38\textwidth]{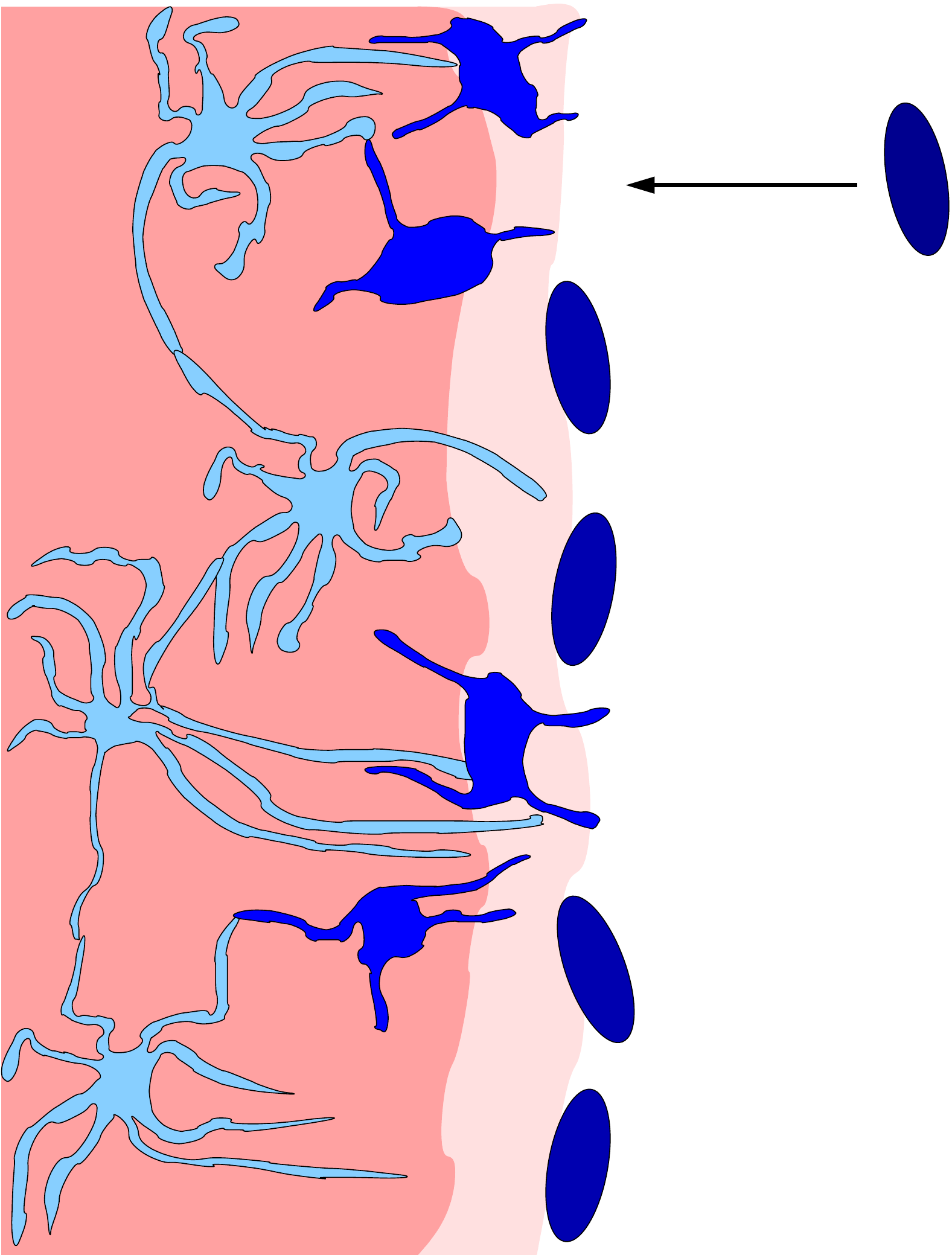} 
   	\put(37.5,93.5){\scriptsize \textcolor{white}{A}} 
	\put(38,40.5){\scriptsize \textcolor{white}{A}} 
	\put(72,85){\scriptsize \textcolor{white}{C}} 
	\put(45,71){\scriptsize \textcolor{white}{B}} 
	\put(45.5,52.5){\scriptsize \textcolor{white}{B}} 
	\put(46.5,22){\scriptsize \textcolor{white}{B}} 
	\put(45,6.5){\scriptsize \textcolor{white}{B}} 
  \end{overpic} 
\caption{\footnotesize{Diagrammatic illustration of the bone-formation process. Lighter shades of blue indicate more differentiated cells. The lighter shade of pink indicates the deposition front, and the darker shade of pink indicates the mineralization front. The top panel occurs earlier than the bottom panel. Dendritic osteocytes (light blue) have dendrites that extend towards the osteoblast layer (dark blue). The osteoblasts secrete bone matrix. Osteoblast cells marked with ``A'' are signaled by the osteocyte network to differentiate into osteocytes. Osteoblast cells marked with ``B'' do not differentiate and stay on the outer bone surface. Osteoblast cells marked with ``C'' arrive at the bone front after differentiating from precursor osteoblasts (pre-osteoblasts). [This figure is inspired by a similar illustration in Ref.~\cite{Franz-Odendaal_2006}.]}}
  \label{fig_bone_laying_diagram}
\end{figure}
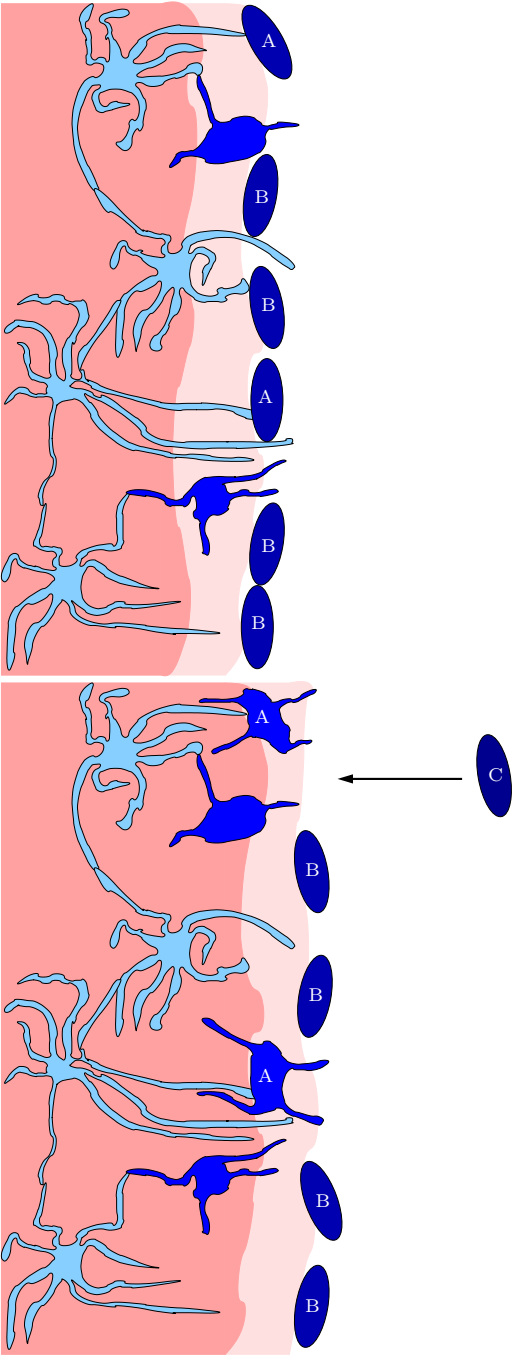

Thus far, we have discussed three types of bone cells: osteoblasts, osteoclasts, and osteocytes. For at least the osteoblast-to-osteocyte cell transition, biologists have subdivided the process of cell differentiation to include eight phenotypes: (\emph{i}) preosteoblast; (\emph{ii}) preosteoblastic osteoblast; (\emph{iii}) osteoblast; (\emph{iv}) osteoblastic osteocyte; (\emph{v}) osteoid-osteocyte (i.e., Type-II preosteocyte); (\emph{vi}) Type-III preosteocyte; (\emph{vii}) young osteocyte; and (\emph{viii}) old osteocyte \cite{Franz-Odendaal_2006}. Additionally, the secretion of bone occurs as two steps: first osteoid is deposited as a scaffold, and then mineralization occurs to confer strength. Stages (\emph{iv})--(\emph{vi}) are cells after the deposition front but before the mineralization front; they are surrounded by a non-mineralized osteoid matrix. (In other words, there is scaffold around them.) Stages (\emph{vii})--(\emph{viii}) are cells whose volume has depleted (by reduction in the endoplasmic reticulum and Golgi apparatus) and are in mineralized bone. The diagram in Fig.~\ref{fig_bone_laying_diagram} shows the bone-formation step. Although it is potentially useful to consider all of the above phases (defined from osteogenic markers), we are interested only in the structure of a mature osteocyte network [stages (\emph{vi})--(\emph{viii})], so we will make drastic simplifications.

%%%%%%%%%%%%%%%%%%%%%%%%%%%%%%%%%%%%

%%%%%%%%%%%%%%%%%%%%%%%%%%%%%%%%%%%%

%%%%%%%%%%%%%%%%%%%%%%%%%%%%%%%%%%%%%%%%%%%%
\subsubsection{Bone Metastasis}
%%%%%%%%%%%%%%%%%%%%%%%%%%%%%%%%%%%%%%%%%%%%

Advanced prostate, breast, and lung cancer can metastasize to bone \cite{Roudier_2003, Zhang_2013}. In pathological bone, the highly regulated bone-remodeling signaling pathway is disrupted. A particularly painful symptom is net bone formation in some regions and simultaneous weakening in other areas \cite{Pivonka_2010}. Small bone lesions can also develop.

In a cancerous microenvironment, transforming growth factor beta (TGF$\beta$) expressed by tumor cells promotes excessive osteoblast growth \cite{Lee_2011}. The overexpression of TGF$\beta$ is only one of the many ways in which tumors can interfere with bone formation; TGF$\beta$ targeted cancer therapy has only been marginally successful \cite{Connolly_2012}. 

Ordinary differential equation (ODE), and hence non-spatial, models for cell populations in healthy
bone that incorporate osteoblast, osteoclast, and osteocyte populations were
developed in \cite{Graham_2013, Moroz_2006, Wimpenny_2007}. Partial differential equation (PDE) models of
healthy bone remodeling include 
\cite{Ryser_2010, Ryser_2009}, and these models were adapted subsequently 
for cancerous bone in \cite{Ryser_2012}.  
Mechanically-focused models that capture stresses and strains in bone
have also been explored \cite{Pivonka_2010, Rejniak_2011,
  Oers_2014}. A few of these models consider osteocyte density, but
none of them explore network structure. 

A seemingly unexplored area is the investigation of osteocyte network
morphology in the presence of cancer. There is evidence in
\cite{Stinson_1975,Eisenberger_2008} that for myeloma and
(benign) osteoma, osteocytes are exceptionally spherical and have
shorter, distorted dendrites that are reduced in number. A contrasting osteocyte network with
unregulated excessive dendritic growth (and hence larger numbers of dendrites) was observed in the presence of osteogenic sarcoma \cite{Stinson_1975}. 

In Section \ref{sec_our_model}, we develop a simple model that incorporates
network properties into the bone-formation process.  
For some cancer types, it is known in part how bone formation is
affected (e.g., there is increased osteoblast proliferation). Using such a model, it
may be possible to connect a change in the bone-formation process to the properties of the resulting
osteocyte network. Qualitatively, one would then 
be able to suggest which osteocyte network phenotype is promoted by a particular cancer (e.g., stunted dendrite growth, excessive dendrite growth, etc.). Conversely it may be possible to infer changes in the bone-formation process from observations of osteocyte network structure.

%%%%%%%%%%%%%%%%%%%%%%%%%%%%%%%%%%%%
\subsection{Model of Osteocyte Network Growth}\label{sec_our_model}
%%%%%%%%%%%%%%%%%%%%%%%%%%%%%%%%%%%%

The model that we develop in this section builds on the work of Buenzli
\cite{Buenzli_2015}.
 We avoid modeling the full complications of the biology (e.g., cell
    sub-classifications, proteins, etc.) and consider only osteoblasts and
    osteocytes. By examining a simple model, we hope to gain insight into how
    osteocyte network structure may depend on measurable
    quantities, while preserving a minimalist approach.

The osteocyte network occupies an expanding domain $\Om(t)\subset\mathds{R}^d$
with a boundary $\p\Om(t)$ that moves with normal velocity $v(t)$. The nodes in the network
are osteocytes, which each have an associated position in $\mathds{R}^d$. The undirected edges are the dendrites between them.  
We allow multiedges, which correspond to multiple dendritic connections between the
same pair of osteocytes, and they are observed in practice
\cite{Kerschnitzki_2013}.  We suppose that the positions of the osteocytes are time-independent, so
$\LO^{(1)} \equiv 0$.

Motivated by the observation that osteoblasts
 differentiate to osteocytes near the mineral front
 \cite{Franz-Odendaal_2006} --- where there is less mineralization --- we suppose that the rate at which an osteocyte creates
 connections to  others is governed by the local bone
 mineral density $\Min(t,\xv)\in\mathds{R}^{+}$. 

Our model consists of two processes: (1) domain expansion and (2) edge
creation within semi-mineralized bone. Buenzli's model
\cite{Buenzli_2015} consists entirely of domain expansion, whereas we
also incorporate a network structure.  

%%%%%%%%%%%%%%%%%%%%%%%%%%%%%%%%%%%%
\subsubsection{Domain Expansion}
%%%%%%%%%%%%%%%%%%%%%%%%%%%%%%%%%%%%

We suppose that osteoblasts are encased within the mineral matrix and
become osteocytes with degree $k=0$ at a rate of
$\Db(t,\xv)\Ob(t,\xv)$ for $\xv \in \p\Om(t)$, where $\Db(t,\xv)$ is the probability
per unit time of an osteoblast 
  joining the matrix and $\Ob(t,\xv)$ is the surface density of
  osteoblasts. (We take $\Ob(t,\xv)$ to be given.)  
Equation (\ref{eq_final_IPDE}) is then
\begin{equation}\label{eq_choice_rj_newph}
	\rj\NewPh(t,\xv)  =
        D_\text{burial}(t,\xv)\Ob(t,\xv)\de_{\p\Om(t)}(\xv)\,, 
\end{equation}
where
\begin{equation*}
	\de_{\p\Om(t)}(\xv) = \int_{\p \Om(t)} \delta(\xv-\xv')\, \dd S'\,.
\end{equation*}	

 Following \cite{Buenzli_2015}, we suppose that the (outward) normal velocity of
the interface is
\begin{equation}\label{eq_boundary_speed}
	v(t,\xv) = \Kf (t,\xv)\Ob(t,\xv)\,, \quad \xv \in \p\Om(t)\,,
\end{equation}
where $\Kf(t,\xv)$ is the volumetric rate at which osteoblasts form the mineral matrix.

%%%%%%%%%%%%%%%%%%%%%%%%%%%%%%%%%%%%
\subsubsection{Edge Creation in Semi-Mineralized Bone}
%%%%%%%%%%%%%%%%%%%%%%%%%%%%%%%%%%%%

We suppose that the mineral density $\Min(t,\xv)$ of the matrix can vary in both space and time, as mineral is produced by both
osteoblasts on the deposition front and osteocytes behind it. 
We suppose that there is a maximum mineral density $\CMin$
that corresponds to fully mineralized bone, that osteocytes produce
mineral at a rate of $\ExcrCy (1 -{\Min}/{\CMin})$
per cell, and that osteoblasts produce
mineral at a rate of $\ExcrOb$
per cell,
where $\ExcrCy$ and $\ExcrOb$ are constants.
Therefore,
\begin{widetext}
\begin{equation}\label{eq_min_density}
	\frac{\p}{\p t} \Min(t,\xv) = \ExcrCy
        \paulF(t,\xv) \left[ 1 -
            \frac{\Min(t,\xv)}{\CMin}\right] +   \ExcrOb \Ob(t,\xv) \de_{\p\Om(t)}(\xv) \, , 
\end{equation}
\end{widetext}
where $\paulF(t,\xv)$ [given by equation (\ref{fandpk})] is the density of osteocytes.
Note that $\Kf$ is the volume of matrix produced per osteoblast per
unit time, whereas $\ExcrOb$ is the mass of matrix produced per osteoblast per
unit time, so the density of matrix formed by the
osteoblasts is $\ExcrOb/\Kf$, which should be no larger than the maximum
mineral density $\CMin$. Note additionally that because $D_\text{burial}$ is the
rate at which osteoblasts join the matrix and $\Kf$ is the rate at
which matrix volume is produced, the ratio $D_\text{burial}/\Kf$ is
the density of newly formed osteocytes.

We now model how edges (i.e., dendrites) form. We suppose that the rate of
edge creation depends on the mineral density $\Min$.
In \cite{Kamioka_2001}, it was suggested that osteocytes grow
dendrites towards the osteoblast layer. This suggests that dendrites
grow in the part of the domain that 
is not fully mineralized. 
To construct a simple model in which edges are less likely to form as mineral density becomes larger, we let the rate of edge creation 
between nodes at positions $\xv$ and $\yv$ be 
\begin{equation}\label{eq_choice_FCon}
	\FCon(t, \xv, \yv) =  (\CMin- \Min (t,\xv))(\CMin -
        \Min(t,\yv))g(||\xv-\yv||)\,,
\end{equation}
where  $g$ is monotonically decreasing and vanishes at infinity,
 encoding the fact that short edges are much more likely to form
than long edges. Equation (\ref{eq_choice_FCon}) takes into account
the mineral density only at the two endpoints $\xv$ and $\yv$. One can formulate more
complicated models in which $\FCon$ 
depends on (for example) a line integral of $\Min$ between $\xv$ and
$\yv$. However, given the simplifications and modeling assumptions that we have already
made, we do not consider such
complicated edge-creation models.

With our model assumptions, equation (\ref{eq_final_IPDE}) becomes
\begin{widetext}
\begin{equation} \label{eq_master}
	\frac{\p}{\p t}u_k(t,\xv) = \left( \int_{\Om(t)} \FCon(t, \xv,\yv) \paulF(t,\yv) \dd \yv \right) \left[ u_{k-1}(t,\xv) - u_k(t,\xv) \right] + \Db(t,\xv)\Ob(t,\xv) \de_{\p\Om(t)}(\xv) \de_{k,0}\,.
\end{equation}
\end{widetext}
Equation \eqref{eq_master} is coupled through $\FCon$ to   
equation (\ref{eq_min_density}) for the mineral density, and the domain
$\Omega(t)$ evolves according to equation (\ref{eq_boundary_speed}).

Summing (\ref{eq_master}) over $k$ yields
\begin{equation}
	\frac{\p \paulF(t,\xv)}{\p t} = \Db(t,\xv)\Ob(t,\xv) \de_{B(t)}(\xv) \, ,
\end{equation}
which is identical to equation (8) in \cite{Buenzli_2015}. Note
also that equation (\ref{eq_master}) does not include feedback between
network structure and the burial rate $\Db$, though one can extend the model to incorporate such coupling.

%%%%%%%%%%%%%%%%%%%%%%%%%%%%%%%%%%%%
\subsection{Traveling-Wave Solution}
%%%%%%%%%%%%%%%%%%%%%%%%%%%%%%%%%%%%

If $\Db$, $\Kf$, and $\Ob$ are constant, then equations
\eqref{eq_boundary_speed}, \eqref{eq_min_density}, and
\eqref{eq_master} admit a 
traveling-wave solution that corresponds to the sustained creation
of new bone. Such a solution may give an indication of the local behavior near growing bone, and we can use it to determine how
the properties of the bone depend on the parameters in the model. 

For a one-dimensional traveling wave that moves with a constant speed of $v = \Kf\Ob $, we can (without loss of generality) take the domain to be \mbox{$\Omega (t) = (-\infty,  vt)$}. We transform to coordinates that move with the wave
by writing $z = x- v t $, and we seek a solution in which $u_k$ and $m$ depend only on $z$. Equation \eqref{eq_master} becomes 
\begin{align}\label{eq_u_k_u}
	-v\frac{\dd}{\dd z}u_k(z) &= \Db\Ob \de (z )\de_{k,0} \\
&+\left( \int_{-\infty}^{0} \FCon(z,z') \paulF(z')\dd z'\right) \left( u_{k-1}(z) - u_k(z) \right) \,.  \nonumber
\end{align}
Summing over $k$ gives
\begin{equation*}
	-v\frac{\dd}{\dd z}f(z) = \Db\Ob \de (z )\,,
\end{equation*}
so 
\begin{equation}\label{eq_varrho_osteocyte_sol}
	\paulF(z) = \frac{\Db}{\Kf}\left[1 - H(z)\right]\,,
\end{equation}
where $H(x)$ is the Heaviside function. Equation \eqref{eq_varrho_osteocyte_sol} is identical to equation
(13) in \cite{Buenzli_2015} when $\Db$ and $\Kf$ are constant. Using (\ref{eq_varrho_osteocyte_sol}) in
equation (\ref{eq_u_k_u}) gives 
\begin{equation}
	\,\frac{\dd}{\dd z}u_k(z) =  -\shA \de
          (z)\de_{k,0}
	-\shB(z)\left( u_{k-1}(z) - u_k(z) \right) \, , \label{eq_u_k_trav_wave}
\end{equation}
where
\begin{equation*} 
	\shA = \frac{\Db}{\Kf}\,, \quad
\shB(z) = \frac{ \Db}{\Kf^2 \Ob}\int_{-\infty}^{0} \FCon(z,z') \dd z'\,.
\end{equation*}
Equivalently, one can write equation \eqref{eq_u_k_trav_wave} as 
\begin{equation}
	\frac{\dd}{\dd z}u_k(z) = -\shB(z) ( u_{k-1}(z) - u_k(z)) \,,
        \quad u_k(0) = \shA\,\de_{k,0}\,,
\end{equation}
whose solution is
\begin{equation}\label{above2}
	u_k(z) =  
\frac{\shA \lambda^k \ee^{-\lambda}}{k!}\,, \quad
	\lambda = \int_z^0 \shB(z' )\dd z'\,. 
\end{equation}
The density of newly-formed osteocytes $b$ sets the scale for $u_k$. 

Before we can solve for $\lambda$, we first need to solve for $\Min(z)$.
Solving equation (\ref{eq_min_density}) yields
\begin{equation}
	\Min(z) = \CMin +  \left( \frac{\ExcrOb}{\Kf}
          -\CMin\right)\exp{ \left( \frac{\ExcrCy \Db z}{ \Kf^2 \Ob
              \CMin } \right) }  \,.
\label{eq_CMin_trav_wave}
\end{equation}
We see that the mineral density
varies from that produced by osteoblasts ${\ExcrOb}/{\Kf}$ to the
maximum mineralization $\CMin$ over a length scale of
\begin{equation*} 	
	L = \frac{ \Kf^2 \Ob
              \CMin }{\ExcrCy \Db }\,.
\end{equation*}
We illustrate this behavior in Fig.~\ref{fig_trav_wave_sol}(top).

Before we can evaluate $\lambda$, we need to choose a form for $g$. 
Suppose first that $g$ is a constant, and let's write $g \equiv \beta$. In principle, 
this may allow long edges to form --- recall (see equation \eqref{eq_choice_FCon}) that we argued that $g$ should decay at infinity to ensure a low probability for long edges to form --- but we see from equation (\ref{eq_CMin_trav_wave}) 
that even with $g$ identically constant, $C(z,z')$ approaches $0$ exponentially
fast over the length scale $L$, which therefore sets the scale for the
maximum edge length.

Inserting equation (\ref{eq_CMin_trav_wave}) in equation (\ref{above2}) gives $ \lambda =
\lambda_{\infty}\left(1-\ee^{z/L}\right)$, where 
\begin{equation}
	 \lambda_\infty
= \frac{ \beta \Ob 
  \CMin^2 \Kf^2}{\Db(\ExcrCy)^2 }\left(   \CMin-\frac{\ExcrOb}{\Kf} 
  	        \right)^2\,, \label{eq_analytic_lambda}
\end{equation}
which indicates that the mean degree varies over the same length scale
from a value of $0$ for newly-formed osteocytes to a value of $\lambda_{\infty}$
deep within a bone.

In the bottom panel of Fig.~\ref{fig_trav_wave_sol}, we illustrate the traveling-wave profile of the degree
distribution by plotting  $(1/b)\sum_{k=0}^K u_k(z)$ for several values of
$K$. The differences between these curves indicate the proportion of
osteocytes of each degree.
As expected, there is  a
  region at the front of the wave in which the mean degree of the osteocytes
  is lower, while the LSDD approaches a
 stationary distribution far behind the front.

\begin{figure}
  \centering
%  \begin{overpic}[width=0.5\textwidth]{trav_wave_Min.eps}
%  	\put(70,2){\large $\omega$}
%	\put(66,41){\large $\Min(z)$} 
%	\put(5,37){\large $\CMin$}
%	\put(2,17){\large $\gamma\CMin$}
%	\put(44,1){\large $z$}  
 % \end{overpic}  \\
\resizebox{0.45\textwidth}{!}{\input{jonFig10a.pdf_t}}\\[5mm]
%    \begin{overpic}[width=0.5\textwidth]{trav_wave_n_k.eps}
%  	\put(70,2){\large $\omega$}
%	\put(66,41){$\sum_{l=0}^k u_l(z)$} 
%	\put(20,39){\small $k\rightarrow\infty$}
%	\put(1,37.25){\small $\paulF = \shA$}
%	\put(1,6.5){\small $k=0$}
%	\put(1,11.5){\small $k=1$}
%	\put(1,19.25){\small $k=2$}
%	\put(1,26.25){\small $k=3$}
%	\put(44,1){\large $z$}   
%  \end{overpic}
\resizebox{0.45\textwidth}{!}{\input{jonFig10b.pdf_t}}\\
  \caption{\footnotesize{(Top) Traveling-wave profile of 
      mineral density $m$ normalized by $\CMin$. At
      the front of the wave, $m=\ExcrOb/\Kf$ (dotted line), which we choose to be $0.5 \CMin$
      in the figure. Behind the front, $m$ approaches $\CMin$. (Bottom) The
      solid curve is a traveling-wave profile of osteocyte density $\paulF= \sum_{k=0}^\infty
      u_k$, normalized by $b = \Db/\Kf$. We
      illustrate the degree distribution of osteocytes as a function
      of position by showing $\sum_{k=0}^K
      u_k$ (dashed curves) for the case $\lambda_{\infty} = 5$. The lower ($K=0$) curve illustrates the proportion of osteocytes
      with degree $0$, and the difference between the $K=i$ and
      $K=i-1$ curves illustrates the proportion of osteocytes with degree $i$.
}} 
  \label{fig_trav_wave_sol}
\end{figure}

If, instead of taking $g$ to be constant, we instead choose $g(z,z') =
\beta \ee^{-|z-z'|/\len}$ or $g(z,z') =
\beta \ee^{-|z-z'|^2/\len^2}$, we obtain qualitatively similar
results, provided $l>L$.
For example, in the first case,
\begin{equation*} 
	\lambda = \lambda_{\infty}\frac{\len (\len - \len\ee^{z/L +
  z/\len}-L + L \ee^{2 z/L}   )}{\len^2-L^2 }\,.
\end{equation*} 
For $\len \gg L$, we see that $\lambda \to \lambda_{\infty}\left(1-\ee^{z/L}\right)$. For $\len \ll L$, we see that $\lambda \to \lambda_{\infty}(\len/L) \left(1-\ee^{2z/L}\right)$, so the
length scale over which degree varies is halved and the mean
degree smaller by the factor $\len/L$.

%%%%%%%

\subsection{Parametrization and Interpretation}

One can infer representative values of some of the parameters and variables in our
model from existing 
experimental data. For example, in the review paper ~\cite{Buenzli_2015_b}, the authors calculated the number of osteocytes per
$\text{mm}^3$ to lie in the range
$19000$--$28500\,\text{mm}^{-3}$; the
surface density $\Ob$ of osteoblasts per $\text{mm}^2$ was calculated
in \cite{Buenzli_2014} (using data from Ref.~\cite{Marotti_1975}) to be in the range
$2000$--$10000\,\text{mm}^{-2}$; and values of $\Db$ and $\Kf$
were given in \cite{Buenzli_2015} (using data from Refs.~\cite{Marotti_1975, Hannah_2010}) for 
bone near a Haversian canal (which occurs only in cortical bone). The number of dendrites that protrude from an
osteocyte yielded a mean degree of
 $\langle k \rangle \approx 52.7$ in \cite{Sugawara_2005}. 
 As suggested in \cite{Kamioka_2001, Sugawara_2011, Kerschnitzki_2013}, we
expect the number of unique neighbors of an osteocyte to be smaller by an
order of magnitude. Measurements of the speed of the mineralization front of bone creation were given in
\cite{Araujo_2014}. Numerous other parameters (e.g., the rate $\ExcrCy$ at which osteocytes produce mineral and the maximum mineral density $\CMin$) are unknown, and the values above were reported in only a few papers, so it is not yet possible to make quantitative predictions with our
model.

Despite the dearth of knowledge about parameter values, we can use our model to make qualitative predictions of
the effect of varying each parameter.
 The review \cite{Logothetis_2005} summarized how different cancer
 types interfere with healthy bone remodeling.
 In particular, the
 review details how factors produced by prostate cancer cells lead to
 net bone formation due to increased levels of proliferation and
differentiation of osteoblasts. In our model, an increased level of
differentiation corresponds to an increased burial rate
$\Db$. Increasing $\Db$ in the model increases the
number density of osteocytes $f$, but it decreases the mean degree $\lambda$. In other words, it decreases the mean number
of dendrites that leave a cell body. In the model, this occurs because
an increase in osteocyte density leads to an increase in
mineralization, which makes dendrite formation less likely.

It was shown in \cite{Stinson_1975,Eisenberger_2008} that for myeloma and benign
osteoma---two other cancers that can cause net bone formation 
\cite{Bataille_1990}---osteocytes are rather spherical with
shorter, distorted dendrites that are fewer in number in comparison to those in healthy tissue. Although these articles do not comment on osteocytes density or on mineralization, and it is crucial to be careful to avoid over-interpreting such limited data, our model
does suggest one possible mechanism for this change in morphology.

%%%%%%%%%%%%%%%%%%%%%%%%%%%%%%%%%%%%%%%%%%%%
%%%%%%%%%%%%%%%%%%%%%%%%%%%%%%%%%%%%%%%%%%%%
%%%%%%%%%%%%%%%%%%%%%%%%%%%%%%%%%%%%%%%%%%%%
%%%%%%%%%%%%%%%%%%%%%%%%%%%%%%%%%%%%%%%%%%%%
%%%%%%%%%%%%%%%%%%%%%%%%%%%%%%%%%%%%%%%%%%%%
%%%%%%%%%%%%%%%%%%%%%%%%%%%%%%%%%%%%%%%%%%%%
%%%%%%%%%%%%%%%%%%%%%%%%%%%%%%%%%%%%%%%%%%%%
%%%%%%%%%%%%%%%%%%%%%%%%%%%%%%%%%%%%%%%%%%%%
%%%%%%%%%%%%%%%%%%%%%%%%%%%%%%%%%%%%%%%%%%%%
%%%%%%%%%%%%%%%%%%%%%%%%%%%%%%%%%%%%%%%%%%%%

%%%%%%%%%%%%%%%%%%%%%%%%%%%%%%%%%%%%%%%%%%%%
\section{Conclusions}\label{conc}
%%%%%%%%%%%%%%%%%%%%%%%%%%%%%%%%%%%%%%%%%%%%

%%%%%%%%%%%%%%%%%%%%%%%%%%%%%%%%%%%%%%%%%%%%
%\subsection{Conclusions and Discussion}
%%%%%%%%%%%%%%%%%%%%%%%%%%%%%%%%%%%%%%%%%%%%

We introduced a model for evolving spatial networks, and we used a mean-field approximation to reduce
the dimension of its governing hierarchal Fokker--Planck equations. Specifically, by defining a local state degree
distribution, we derived IPDEs \eqref{eq_final_IPDE} and \eqref{eq_del_addition} to
describe an evolving spatial network that includes
evolution of the position of nodes (or some more general state vector), 
edge creation, edge deletion, and new node creation
that occur at prescribed rates. Our approach generalizes commonly-studied master-equation approaches by including a state space so that we can examine spatial networks. 
 
To illustrate the potential utility of our IPDEs in applications, we examined growing osteocyte networks in bone. Although we employed a very simplistic model, we were able to use it predict relationships between biological parameters and network structure. Our approach provides a  starting point for examining spatial networks in biology and other fields. In the future,
we hope to compare predictions to experimental data after making a
model more faithful to the biology.

%%%%%%%%%%%%%%%%%%%%%%%%%%%%%%%%%%%%%%%%%%%%
\section{Acknowledgements}
%%%%%%%%%%%%%%%%%%%%%%%%%%%%%%%%%%%%%%%%%%%%

J. P. T.-K. received funding from the EPSRC under grant reference number EP/G037280/1. We thank Pascal Buenzsli, Mikko Kivel{\"{a}}, Andrew Krause, SeWook Oh, Jan Poleszczuk, and particularly Paul J. Dellar for helpful discussions.

%%%%%%%%%%%%%%%%%%%%%%%%%%%%%%%%%%%%%%%%%%%%
%%%%%%%%%%%%%%%%    REFERENCES     %%%%%%%%%%%%%%%%%%
%%%%%%%%%%%%%%%%%%%%%%%%%%%%%%%%%%%%%%%%%%%%

%\bibliographystyle{siam}
%\bibliography{References3.bib}

%%%%%%%%%%%%%%%%%%%%%%%%%%%%%%%%%%%%%%%%%%%%
%%%%%%%%%%%%%%%%    REFERENCES     %%%%%%%%%%%%%%%%%%
%%%%%%%%%%%%%%%%%%%%%%%%%%%%%%%%%%%%%%%%%%%%

%%%%%%%%%%%%%%%%%%%%%%%%%%%%%%%%%%%%%%%%%%%%
%%%%%%%%%%%%%%%%       APPENDIX       %%%%%%%%%%%%%%%%%%
%%%%%%%%%%%%%%%%%%%%%%%%%%%%%%%%%%%%%%%%%%%%
\appendix %\small
%%%%%%%%%%%%%%%%%%%%%%%%%%%%%%%%%%%%%%%%%%%%
%%%%%%%%%%%%%%%%       APPENDIX       %%%%%%%%%%%%%%%%%%
%%%%%%%%%%%%%%%%%%%%%%%%%%%%%%%%%%%%%%%%%%%%

%%%%%%%%%%%%%%%%%%%%%%%%%%%%%%%%%%%%%%%%%%%%
\section{Candidate Algorithm for the Kinetic Network
  Model}\label{App_algorithm} 
%%%%%%%%%%%%%%%%%%%%%%%%%%%%%%%%%%%%%%%%%%%%

In Algorithm \ref{Algo_network_generation}, we give pseudocode for our
simulations of evolving spatial networks. In the main text, we summarized our model in Table \ref{table_full_model}. 

When simulating Algorithm \ref{Algo_network_generation}, we use a
small time step $\De t$, so $0< \De t \ll 1$. We also
specify the following ordering of events: edge creation; edge deletion; state
update; and then new particles are allowed to enter the system.   
This specification is arbitrary, and obviously it is desirable that any reordering of these events becomes
inconsequential as $\De t \rightarrow 0$. For our numerical experimentation using the examples in Section \ref{sec_numerics}, this indeed appears to be the case. For the simulations that we reported in Section \ref{sec_numerics}, we chose a time step $\De t$ to
be sufficiently small that a reordering of events has no discernible impact.

The update rule for event (\emph{iv}) in Table
\ref{table_full_model}---namely, the evolution of the states of the
individual nodes---depends on the particular process that we consider. 
We write \mbox{$\sv_i(t + \De t) =
\mathscr{D}(\sv_1(t),\dots,\sv_{N(t)}(t), \De t)$}, where $\mathscr{D}$
arises from the time-discretization of the underlying
process.  For example, one can use an Euler--Maruyama method or the Milstein method for the SDE \eqref{eq_SDE_EoM}
\cite{Higham_2001}; and one use St\"{o}rmer--Verlet schemes for the ODEs in equation
\eqref{eq_Newton_EoM}.

%%%%%%%%%%%%%%%%%%%%%%%%%%%%%%%%%%%%
% ALGORITHM 1
%%%%%%%%%%%%%%%%%%%%%%%%%%%%%%%%%%%%
\begin{widetext}
\small
\begin{algorithm}[H]
\SetCommentSty{emph}
%\SetKwComment{comment}{\%}{}
\DontPrintSemicolon
 \KwData{Choose an end time $T_{\text{end}} = M\De t$ for large $M\in\mathds{N}$ and small $\De t >0$.\;
 Set the number of particles $N \leftarrow N_0$.\; Initialize the
 starting state at 
 $\sv_{i} \leftarrow \sv_0^{(i)}$ and starting degree at $k_i \leftarrow
 k_0^{(i)}$ for each $i \in \{1,\dots,N_0\}$.} 
 Set time counter $m\leftarrow 0$.\;
 \While{$m \leq M$}{
 \comment{Edge creation update.}
  \For{$i\leftarrow$ \texttt{randperm}$(\{1,2,\dots,N\})$}{
  	\For{$j\leftarrow$ \texttt{randperm}$(\{i+1,\dots,N\})$}{
  Draw a uniform random number $r_1$ from the distribution $\mathcal{U}(0,1)$.\;
  \If{$r_1 \leq \FCon (\sv_j,k_j|\sv_i,k_i)\De t $}{
  Create an edge between node $i$ and node $j$.
  \;}
  }
  }
  \comment{Edge deletion update.}
  \For{$i\leftarrow$ \texttt{randperm}$(\{1,2,\dots,N\})$}{
  	\For(\comment*[h]{$E_i$ is the set nodes joined by
            an edge to $i$, counted according to multiplicity.}){$j\leftarrow$ \texttt{randperm}$(E_i)$}{
  Draw a uniform random number $r_2$ from the distribution $\mathcal{U}(0,1)$.\;
  \If{$r_2 \leq \FDel (\sv_j,k_j |\sv_i,k_i) \De t$}{
  Delete an edge between node $i$ and node $j$.
  \;}
  }}
  \comment{State update.}
  \For{$i\leftarrow 1$ \KwTo $N$}{
  Update particle state: $\sv_i \leftarrow \mathscr{D}(\sv_1,\dots,
  \sv_N, \De t)$. \comment{$\mathscr{D}$ arises from the
    time-discretization of the state dynamics} 
  }
  \comment{New node creation.}
  Draw a uniform random number $r_3$ from the distribution $\mathcal{U}(0,1)$.\;
  \If{$r_3 \leq \rj\De t$}{
  Create node with state $\sv_{N + 1} \leftarrow \sv^*$,
  where $\sv^* \sim \NewPh $. \;
  Initialize the degree $k_{N + 1} \leftarrow 0$ \; 
  Update the number of particles: $N \leftarrow N +1 $.}
  Update time: $m \leftarrow m +1 $.
  }
 \caption{\footnotesize{Algorithm to generate an evolving spatial
     network. The notation
     \texttt{randperm}$(X)$ signifies a permutation, selected uniformly at random, of the discrete set
     $X$. 
 }}
\label{Algo_network_generation}
\end{algorithm}
\normalsize
\end{widetext}
%%%%%%%%%%%%%%%%%%%%%%%%%%%%%%%%%%%%

One can devise a efficient simulation algorithm for
situations in which edge creation or deletion do not depend on the state
of nodes. In that case, one can use an event-driven algorithm, such as a Gillespie
algorithm \cite{Gillespie_1977}, for creation and deletion events.

\section{Additional Details of the Model
 Derivation: No Edge 
  Deletion}\label{App_finish_derivation_creation} 
%%%%%%%%%%%%%%%%%%%%%%%%%%%%%%%%%%%%%%%%%%%%

In this appendix, we fill in the details of the derivation in Section
\ref{lowdim}. 
When we sum over the degrees and integrate over the states of particles 2 through
$n$, multiply by $n$, and sum over $n$, the 
 first term on the RHS of equation (\ref{eq_HFP_l1}) gives
\begin{widetext}
\begin{align}
% Original 
	&\,\sum_{n=0}^\infty n  \shSUMk{2}{n}   \int_{\mS^{n-1} } \sum_{i=1}^n \sum_{j
          =i+1}^n   \FCon(\sv_i, k_i-1, \sv_j, k_j-1)
          \FF_n^{\,\kv_{n,-}^{ij} }(t, \vs_n)   \, \dd
          \vs_{n}^{\,(2)} \nonumber
  \\  	&\qquad \qquad\mbox{ }\label{eq_app_l1}
- \sum_{n=0}^\infty n  \shSUMk{2}{n}  \int_{\mS^{n-1} } \sum_{i=1}^n
          \sum_{j =i+1}^n\FCon(\sv_i, k_i, \sv_j, k_j)
          \FF_n^{\,\kv_n}(t, \vs_n)  \, \dd \vs_{n}^{\,(2)} 
         \, . 
  \end{align}
%\end{widetext}
For $i>1$, each individual term appears once in the positive sum and
once in the negative sum; they thus cancel each other out. 
The remaining terms
are
%\begin{widetext}
\begin{align}
% Original 
	&\, \sum_{n=0}^\infty n \shSUMk{2}{n}   \int_{\mS^{n-1} }  \sum_{j
          =2}^n   \FCon(\sv_1, k_1-1, \sv_j, k_j-1)
          \FF_n^{\,\kv_{n,-}^{ij} }(t, \vs_n)   \, \dd
          \vs_{n}^{\,(2)}  \nonumber
  \\  	&\qquad\qquad \mbox{ }
- \sum_{n=0}^\infty n  \shSUMk{2}{n}  \int_{\mS^{n-1} }
          \sum_{j =2}^n\FCon(\sv_1, k_1, \sv_j, k_j)
          \FF_n^{\,\kv_n}(t, \vs_n)  \, \dd \vs_{n}^{\,(2)} 
        \, . \label{eq_app_l2}
  \end{align}
% \end{widetext}
Because $\FF_n^{\,\kv_n}(t, \vs_n)$ is invariant with respect to
 particle relabeling, we can relabel $(\sv_j, k_j) \leftrightarrow
 (\sv_2, k_2)$ in each term in the sum over $j$ to obtain  
%\begin{widetext}
\begin{align*}
\lefteqn{\sum_{n=0}^\infty n\shSUMk{2}{n}  \int_{\mS^{n-1} }
          \sum_{j =2}^n\FCon(\sv_1, k_1, \sv_j, k_j)
          \FF_n^{\,\kv_n}(t, \vs_n)  \, \dd \vs_{n}^{\,(2)}  = 
\sum_{n=0}^\infty n\shSUMk{2}{n}  \int_{\mS^{n-1} }
          \sum_{j =2}^n\FCon(\sv_1, k_1, \sv_2, k_2)
          \FF_n^{\,\kv_n}(t, \vs_n)  \, \dd \vs_{n}^{\,(2)}}\qquad & \hspace{17cm} \\
 &  =\sum_{n=0}^\infty n (n-1) \sum_{k_2 = 0}^\infty  \int_{\mS } 
         \FCon(\sv_1, k_1, \sv_2, k_2)
         \shSUMk{3}{n}  \int_{\mS^{n-2} } \FF_n^{\,\kv_n}(t, \vs_n)
  \, \dd \vs_{n}^{\,(3)}\, \dd \sv_{2} \\ &
  = \sum_{k_2 = 0}^\infty  \int_{\mS }  
         \FCon(\sv_1, k_1, \sv_2, k_2)
                                            u_{k_1,k_2}^{(2)}(t,\sv_1,\sv_2)
                                            \, \dd \sv_{2}  
\end{align*}
where the last line follows from equation (\ref{2particleULSDD}). Consequently, we can write equation (\ref{eq_app_l1}) as
% \begin{widetext}
\begin{align}
	&\int_{\mS} \sum_{k_2=0}^\infty   \FCon(\sv_1, k_1-1, \sv_2, k_2-1) u_{k_1-1,k_2-1}^{(2)}(t,\sv_1,\sv_2)  \, \dd \sv_{2} 
 - \int_{\mS} \sum_{k_2=0}^\infty  \FCon(\sv_1, k_1, \sv_2, k_2) u_{k_1,k_2}^{(2)}(t,\sv_1,\sv_2)  \, \dd \sv_{2} \label{eq_creation_identify_n_2}  \,  .
\end{align}
% \end{widetext}
For the remaining terms on the RHS of equation (\ref{eq_HFP_l1}), 
we sum over the degrees and integrate over the states of particles 2 through
$n$, multiply by $n$, and sum over $n$ to obtain
% \begin{widetext}
\begin{equation}
	\sum_{n=1}^\infty n  \shSUMk{2}{n}   \int_{\mS^{n-1} } \left(\sum_{i=1}^n \frac{1}{n} \de_{k_i, 0} \,\rj \,\NewPh(\sv_i)
  \FF_{n-1}^{\,\kv_{n}^{i-}}(t,  
\vs_{n}^{i-}) - \rj \FF_n^{\,\kv_n}(t, \vs_n)\right) \, \dd
          \vs_{n}^{\,(2)}\,.\label{source}
\end{equation}
% \end{widetext}
For the term  $i=1$, we are summing and integrating over all 
arguments of $ \FF_{n-1}^{\,\kv_{n}^{i-}}(t,  \vs_{n}^{i-})$, so
% \begin{widetext}
\begin{align*} 
	\sum_{n=1}^\infty n  \shSUMk{2}{n}   \int_{\mS^{n-1} } \frac{1}{n} \de_{k_1, 0} \,\rj \,\NewPh(\sv_1)
  \FF_{n-1}^{\,\kv_{n}^{1-}}(t,  
\vs_{n}^{1-}) \, \dd
          \vs_{n}^{\,(2)}& = \de_{k_1, 0}  \,\rj \,\NewPh(\sv_1)\sum_{n=1}^\infty   \shSUMk{2}{n}   \int_{\mS^{n-1} } \FF_{n-1}^{\,\kv_{n}^{1-}}(t,  
\vs_{n}^{1-}) \, \dd
          \vs_{n}^{\,(2)} \\ & = \de_{k_1, 0}  \,\rj
                               \,\NewPh(\sv_1)\sum_{n=1}^\infty 
                               \shSUMk{1}{n-1}   \int_{\mS^{n-1} }
                               \FF_{n-1}^{\,\kv_{n-1}}(t,   
\vs_{n-1}) \, \dd
          \vs_{n-1}  \\ & = \de_{k_1, 0}  \,\rj \,\NewPh(\sv_1)\,,
\end{align*}
% \end{widetext}
where the last line follows from equation (\ref{eq_FF_normalisation}).
For each term $i>1$, we use the invariance of $\FF_n^{\,\kv_n}(t,\vs_n)$ with respect to particle relabeling to swap particle $i$ with
particle $n$ to give
% \begin{widetext}
\begin{align*}
	\lefteqn{\sum_{n=1}^\infty n  \shSUMk{2}{n}   \int_{\mS^{n-1} }\sum_{i=2}^n
  \frac{1}{n} \de_{k_i, 0} \,\rj \,\NewPh(\sv_i) 
  \FF_{n-1}^{\,\kv_{n}^{i-}}(t,  
\vs_{n}^{i-})  \, \dd
          \vs_{n}^{\,(2)} = \sum_{n=1}^\infty n  \shSUMk{2}{n}   \int_{\mS^{n-1} }\sum_{i=2}^n
  \frac{1}{n} \de_{k_n, 0} \,\rj \,\NewPh(\sv_n) 
  \FF_{n-1}^{\,\kv_{n}^{n-}}(t,  
\vs_{n}^{n-})  \, \dd
          \vs_{n}^{\,(2)} }\hspace{2cm} & \hspace{16cm}\\ 
&=\rj \,    \sum_{n=1}^\infty
               (n-1)
                                        \int_{\mS }
   \,\NewPh(\sv_n) \, \dd\sv_n   \shSUMk{1}{n-1}  \int_{\mS^{n-2} } 
  \FF_{n-1}^{\,\kv_{n-1}}(t,  
\vs_{n-1})  \, \dd
          \vs_{n-1}^{\,(2)} \\ 
&=\rj \,    \sum_{n=0}^\infty
               n     \shSUMk{1}{n}  \int_{\mS^{n-1} } 
  \FF_{n}^{\,\kv_{n}}(t,  
\vs_{n})  \, \dd
          \vs_{n}^{\,(2)} 
\,,
\end{align*}
 \end{widetext}
which cancels with the remaining term in equation (\ref{source}).

%%%%%%%%%%%%%%%%%%%%%%%%%%%%%%%%%%%%%%%%%%%%
\section{Additional Details of the Model Derivation: Edge
  Deletion}\label{App_finish_derivation_deletion}  
%%%%%%%%%%%%%%%%%%%%%%%%%%%%%%%%%%%%%%%%%%%%

In kinetic theory, the first question to address when deriving a reduced model is which
variables to retain in the model and which to integrate over. In Section \ref{sec_edge_creation}, each variable was associated with a node in a network, and it was natural to integrate over all nodes but the first. We thus retained the
state and degree of node 1 as independent variables. We could have
reduced the model further by subsequently integrating over either the state
or degree of node 1.

When considering edge deletion, it is much more difficult to associate the independent variables with individual nodes, and each entry of an adjacency matrix is associated with a pair of nodes.
Consequently, it is not obvious which variables are natural to retain in a
reduced model and which variables should be integrated out.
To facilitate a direct comparison of the reduced model including edge deletion with the reduced model of
Section \ref{sec_edge_creation}, we again retain the state and degree
of node 1 as independent variables. We thus sum over
all entries of the adjacency matrix for which node 1 has degree $k_1$.
Additionally, as before, we integrate over $\sv_n^{(2)}$, multiply by
$n$, and sum over $n$. 

Because the operators $\LO^{(n)}$ that we are considering do not
depend on network structure, the approximation of the LHS of equation \eqref{eq_full_HFP} proceeds as in Section \ref{lowdim}. 
For the edge-creation term on the RHS of equation (\ref{eq_full_HFP})
we find, as in Section \ref{App_finish_derivation_creation}, that for $i>1$,
each term appears once in the 
positive sum and once in the negative sum; these terms thus cancel. For the remaining terms (for which $i=1$), we exploit the invariance of $\FAd_n^{\,\Ad_n}(t, \vs_n)$ with respect to 
 particle relabeling. Specifically, we relabel $j\leftrightarrow 2$ in each
 term in the sum over $j$ (i.e., swapping rows and columns of $A_n$) to obtain 
\begin{widetext}
\begin{align}
	\lefteqn{ \sum_{n=0}^{\infty} n\, 
 \shSUMAdjA{n}{2} \int_{\mS^{n-1} }  \de\left( k_1 ,\, \sum_{j=2}^n
  (\Ad_n)_{1j}   \right) (n-1) \FCon(\sv_1, k_1-1, \sv_2, k_2-1) 
  \FAd_n^{ \Ad_{n,-}^{12} } (t, \vs_n)  
    \, \dd \vs_n^{\, (2)}} \hspace{3cm} & \nonumber\\
&
- \sum_{n=0}^{\infty} n\, 
 \shSUMAdjA{n}{2} \int_{\mS^{n-1} } \de\left( k_1 ,\, \sum_{j=2}^n
  (\Ad_n)_{1j}   \right)  (n-1)   \FCon(\sv_1, k_1, \sv_2, k_2) 
  \FAd_n^{ \Ad_{n} }  (t, \vs_n)   \, \dd \vs_n^{\, (2)}\,,\label{Cdeletion}
\end{align}
%\end{widetext}
where the Kronecker $\delta$ enforces the degree condition. The 2-particle LSDD is
%\begin{widetext}
\begin{equation} \label{eq_def_n_q_Ad} 
	u_{k_1,k_2}^{(2)}(t, \sv_1, \sv_2) = \sum_{n=0}^{\infty} n(n-1)
       \shSUMAdjA{n}{2}   \de\left( k_1
              ,\, \sum_{j=2}^n (\Ad_n)_{1j}   \right)\de\left( k_2
              ,\, (A_n)_{12}+\sum_{j=3}^n (\Ad_n)_{2j}   \right)
            \int_{\mS^{n-2} } \FAd_n^{ \Ad_{n} } (t,
            \vs_n)\, \dd \vs_n^{\, (3)}   \,  ,
\end{equation}
%\end{widetext}
so one can write equation (\ref{Cdeletion}) as equation (\ref{eq_creation_identify_n_2}),
and the analysis proceeds as in Section \ref{App_finish_derivation_creation}. 

Let's now consider the edge-deletion terms.
As with the edge-creation terms, for $i>1$,
each term appears once in the 
positive sum and once in the negative sum; these terms thus cancel each other.
For the remaining terms (for which $i = 1$), relabeling $j\leftrightarrow 2$ in each
 term in the sum over $j$ yields 
%\begin{widetext}
\begin{align}
% Original
 \lefteqn{\sum_{n=0}^{\infty} n (n-1)  \shSUMAdjA{n}{2}\de\left(
               k_1 ,\, \sum_{j=2}^n (\Ad_n)_{1j}   \right)
               \int_{\mS^{n-1} }  
               [(\Ad_n)_{12} + 1]  \FDel(\sv_1, k_1+1, \sv_2, k_2+1)
               \FAd_n^{\Ad_{n,+}^{12} }(t, \vs_n)   \, \dd
               \vs_{n}^{\,(2)} } \hspace{2cm} & \nonumber \\ 
 &\mbox{ }\quad 	-\sum_{n=0}^{\infty} n(n-1)  \shSUMAdjA{n}{2} \de\left( k_1
           ,\, \sum_{j=2}^n (\Ad_n)_{1j}   \right)
           \int_{\mS^{n-1} }  
           (\Ad_{n})_{12} \FDel(\sv_1, k_1, \sv_2, k_2)
           \FAd_n^{\Ad_n}(t, \vs_n)   \, \dd \vs_{n}^{\,(2)}
           \nonumber\\
           & =
           \sum_{k_2=0}^\infty \int_{\mS}  \FDel(\sv_1, k_1+1, \sv_2,
           k_2+1)	\newu_{k_1+1,k_2+1}^{(2)}(t, \sv_1,\sv_2)\,  \dd \sv_{s}- 
           \sum_{k_2=0}^\infty \int_{\mS}  \FDel(\sv_1, k_1, \sv_2,
           k_2)	\newu_{k_1,k_2}^{(2)}(t, \sv_1,\sv_2)\,  \dd \sv_{s} \,,
\end{align}
where
\begin{equation} \label{eq_def_n_2_Ad} 
	\newu_{k_1,k_2}^{(2)}(t, \sv_1, \sv_2) = \sum_{n=0}^{\infty} n(n-1)
       \shSUMAdjA{n}{2}   \de\left( k_1
              ,\, \sum_{j=2}^n (\Ad_n)_{1j}   \right)\de\left( k_2
              ,\, (A_n)_{12}+\sum_{j=3}^n (\Ad_n)_{2j}   \right)
            \int_{\mS^{n-2} }(A_n)_{12} \FAd_n^{ \Ad_{n} } (t,
            \vs_n)\, \dd \vs_n^{\, (3)}   \,  .
\end{equation}
\end{widetext}
There is now a new closure problem, as we need to relate
$\newu_{k_1,k_2}^{(2)}(t, \sv_1, \sv_2)$ to known variables.
We write
\begin{equation*} 
	\newu_{k_1,k_2}^{(2)}(t, \sv_1, \sv_2) = \alpha u_{k_1,k_2}^{(2)}(t, \sv_1, \sv_2)\,,
\end{equation*}	
where $\alpha$ is the expected number of edges between
nodes 1 and 2, given that these nodes have degrees $k_1$ and $k_2$, respectively. If there are
$m$ edges in total, there are $2m$ stubs, of which $k_1$ are at
node 1 and $k_2$ at node 2. One can approximate the probability that a given edge
connects nodes 1 and 2 as 
\begin{equation*} 
	2 \times \frac{k_1}{2m} \times \frac{k_2}{2m}\,,
\end{equation*}	
so a reasonable closure assumption (reminiscent of a configuration model and hence with similar associated assumptions \cite{Fosdick2016}) for the expected number of edges between nodes 1 and 2 is
\begin{equation*} 
	\alpha \approx \frac{k_1 k_2}{2m}  = \frac{k_1k_2}{E[N] \langle k
  \rangle} = \frac{k_1k_2}{\int_{\mS} \sum_{k_1=0}^\infty k_1
  u_{k_1}(t,\sv_1)\, \dd \sv_1} \,,
\end{equation*}  
where $E[N]$ is the expected number of nodes and $\langle k \rangle$
is the mean degree. Using a mean-field approximation for $u_{k_1,k_2}^{(2)}(t,
\sv_1, \sv_2)$, we can then close the edge deletion term by writing
\begin{equation*}
	\newu_{k_1,k_2}^{(2)}(t, \sv_1, \sv_2) = \frac{k_1 k_2 u_{k_1}(t,
\sv_1) u_{k_2}(t, \sv_2)}{\int_{\mS}
  \sum_{k_1=0}^\infty k_1 
 	 u_{k_1}(t,\sv_1)\, \dd \sv_1}\,.
\end{equation*}

Finally, we consider the node-creation term.
The term $i=1$ gives
\begin{widetext}
\begin{align*} 
	\lefteqn{\sum_{n=0}^{\infty} n\, 
 \shSUMAdjA{n}{2} \int_{\mS^{n-1} }  \de\left( k_1 ,\, \sum_{j=2}^n
 (\Ad_n)_{1j}   \right)
   \frac{1}{n} \left[\prod_{j=2}^{n} \de ({0, 
  (\Ad_{n})_{1j}} ) \right] \,\rj \NewPh(\sv_1) \FAd_{n-1}^{ \Ad_{n}^{1-}
   }(t, \vs_{n}^{\,1-}) \, \dd
   \vs_{n}^{\,(2)}
} \hspace{5cm}
   &
\\ & = \de_{k_1, 0}  \,\rj \,\NewPh(\sv_1)\sum_{n=1}^\infty
 \sum_{A_n^{1-} \in S_{n-1}}   \int_{\mS^{n-1} }  \FAd_{n-1}^{ \Ad_{n}^{1-}
   }(t,  
 \vs_{n-1}) \, \dd
	 \vs_{n-1}    = \de_{k_1, 0}  \,\rj \,\NewPh(\sv_1)\,,
\end{align*}
\end{widetext}
where the last equality follows by equation (\ref{deletenormal}).
For each term with $i>1$, we can use the invariance of $\FF_n^{\,\kv_n}(t,
\vs_n)$ with respect to particle relabeling to swap particle $i$ with
particle $n$. As in our prior calculations, we then find that all of these terms cancel each other out.

%%%%%%%%%%%

%%%%%%%%%%%%%%%%%%%%%%%%%%%%%%%%%%%%%%%%%%%%
\end{document}